\documentclass[12pt]{article}

\usepackage{etoolbox}

\newtoggle{unnumberedsections}
\settoggle{unnumberedsections}{false}

\relax

\usepackage{amssymb,amsmath,amsthm}
\usepackage[colorlinks,citecolor=blue,urlcolor=blue]{hyperref}

\usepackage{cases}
\usepackage{graphicx}
\usepackage{epstopdf}
\usepackage{algorithm}
\usepackage[noend]{algpseudocode} 
\usepackage[round]{natbib}
\usepackage{bibunits}
\defaultbibliographystyle{apalike}
\defaultbibliography{Other/Refs}

\usepackage{color}
\usepackage{xcolor}%
\usepackage[dvipsnames]{xcolor}

\usepackage{subcaption} 

\usepackage[toc,page,title,titletoc,header]{appendix}
\usepackage{multirow}
\usepackage{booktabs}

\usepackage{setspace}
\usepackage{lineno} 

\usepackage{url}
 
\usepackage[overload]{empheq}
\usepackage{cleveref}

\usepackage[normalem]{ulem}%
 
\usepackage{xr}

\usepackage[top=2.7cm, bottom=3cm, left=2.7cm, right=2.7cm]{geometry}

\usepackage{mathtools} %

\usepackage{subcaption}
\usepackage{tabularx}
\usepackage{booktabs}

\usepackage{sectsty}%
\usepackage[subfigure]{tocloft}%
\usepackage{titlesec}%

\usepackage{multirow}

\newtheorem{innerremark}{Remark}
\newtheorem{proposition}{Proposition}

\usepackage{amsmath}               
  {
      \theoremstyle{plain}

  }

\usepackage{bbm}
\usepackage{bm}

\usepackage{authblk} 

\usepackage{booktabs}

\title{A generalized Bayesian method for identifying local and global changepoints in dynamic network data}

\begin{document}

\author[]{Yuhui Wang}
\author[]{Michael Jauch\thanks{Email: mjauch@fsu.edu}} 
\author[]{Joshua Daniel Loyal}

\affil[]{Department of Statistics, Florida State University, \\
117 North Woodward Avenue, Tallahassee, Florida 32306, U.S.A.}

    \maketitle


\begin{bibunit}
    

\newcommand{\jl}[1]{\textcolor{orange}{\bf [Josh: #1]}}

    
\begin{abstract} 
We introduce a generalized Bayesian approach to multiple changepoint analysis in dynamic network data based on a multinomial log loss function and a prior distribution that imposes low-rank structure and sparsity on the model parameters. The method can detect local and global changes in network structure by incorporating information from both the full adjacency matrices and network summary statistics. Moreover, the proposed method is agnostic to the true data-generating process. From the generalized posterior distribution, we can make simultaneous inferences with meaningful uncertainty quantification about the locations of changepoints as well as deduce the network attributes and summary statistics that characterize them. For posterior simulation, we develop an efficient Gibbs sampler based on P\'olya-Gamma data augmentation with a computational cost that grows linearly with the number of observed edges in the dynamic network. Simulations across several benchmark network models show that the proposed method is often more accurate and flexible than competitors. We demonstrate the practical value of the method through applications to an email communication network and an international conflict network.


\end{abstract}


\section{Introduction}\label{Sec. Introduction}



Network changepoint analysis, that is, the identification of abrupt structural changes in a time series of networks, is an important task across scientific disciplines. For example, neuroscientists have used network changepoint analysis to identify changes in the brain’s network of neural connections that might indicate the progression of a neurological disease such as Alzheimer's or epilepsy~\citep{khambhati2015dynamic, sengupta2025statistical}. Sociologists have applied network changepoint methods to organizational networks to identify changes in communication patterns induced by the shift to remote work during the COVID-19 pandemic~\citep{zuzul2024}. In each of these cases, the data is a collection of time-indexed networks on $n$ nodes called a dynamic network, where each  network describes the relations, or edges, between pairs of entities, or nodes. The individual networks are often represented by $n \times n$ adjacency matrices with entries that quantify the strength of the relation.

While offline changepoint analysis for vector-valued data has a rich history~\citep{page1954, truong2020}, offline network changepoint analysis is relatively new.  Existing statistical approaches to offline network changepoint analysis can be split into four rough categories: model-based, vectorized, metric space, and embedding methods. Model-based methods assume certain generative models for the networks and look for abrupt changes in the model parameters~\citep{peel2015, corneli2018, bhattacharjee2020, park2020, kei2023cpdstergm}. Vectorized methods convert the dynamic network into a time series of vectors and apply traditional vector-based changepoint localization methods~\citep{priebe2005scan, wang2014, bhattacharyya2020, wang2021cpdnbs, chen2024}. Similarly, metric space methods treat each network as a random object in a general metric space and detect abrupt changes in kernel~\citep{song2024practical} or graph-based~\citep{chen2015, chu2019} dissimilarity measures or in the sequence's Fr\'echet means and variances~\citep{dubey2020}. Although vectorized and metric space methods are often nonparametric and thus model-agnostic, they lose detection power because they ignore information about network structure. As such, recent work has focused on embedding methods, which embed the network's nodes into a low-dimensional latent space and apply changepoint methods to these time-varying node embeddings~\citep{madridpadilla2022cpdrdpg, zhang2024, chen2025, kei2025decoder, athreya2025, chen2026}. For a more nuanced review, see \citet{zhou2025}.

Dynamic network data's complex structure, including strong temporal and cross-sectional dyad dependencies, makes the development of principled statistical methodology for network changepoint analysis difficult. In particular, strong dyad dependencies make model-based approaches~\citep{peel2015, corneli2018, bhattacharjee2020, park2020, kei2023cpdstergm} brittle because of likely model misspecification. In addition, a network's complex structure means changepoints often fall into two distinct types: {\it changepoints in global network structure}, meaning shifts in the stochastic tendencies toward certain local edge configurations across the entire network  (e.g., overall edge density, node degrees, triangle count), and {\it changepoints in local network structure}, meaning shifts in the appearance of local regions in the network such as a shift in the behavior of a small group of nodes (e.g., brain regions affected by a neurological disease). Although there are many methods focused on detecting either global changes~\citep{chen2015, chu2019, zhao2019, bhattacharyya2020, dubey2020, wang2021cpdnbs, chen2024, song2024practical} or local changes~\citep{priebe2005scan, zhao2019,madridpadilla2022cpdrdpg, zhang2024, chen2025, kei2025decoder, athreya2025, chen2026}, there has been less work devoted to developing interpretable methods that can clearly distinguish between the two types of changes. Moreover, very few network changepoint methods provide meaningful uncertainty quantification across the multiple related inferences that arise in changepoint analysis, e.g., inferences regarding the number, locations, and nature of the changepoints.


In this work, we present a generalized Bayesian approach~\citep{bissiri2016general} to network changepoint analysis that satisfies the four desiderata for an offline network changepoint method outlined in the discussion above: The method (1) detects changes in either global or local network structure, (2) provides interpretable summaries that help to distinguish between these two types of changes, (3) offers meaningful uncertainty quantification across related inferences, and (4) is agnostic to the true data-generating process. Although existing approaches address some of these desiderata, they rarely address all four. Moreover, unlike existing Bayesian methods that take a model-based approach~\citep{peel2015, park2020, casarin2025}, our method provides a coherent basis for updating prior beliefs and quantifying uncertainty without relying on a likely misspecified parametric network model. To do so, we build upon the work of \citet{thomas2024bayesian} and \citet{wang2026bcmlr}, which developed a generalized Bayesian methodology for changepoint analysis in vector-valued time series. However, the latter article's approach of converting dynamic networks into time series of vector summaries often fails to detect local changepoints because it ignores information in network structure, motivating our new methodology.

Under the generalized Bayesian framework, our first contribution is the development of a loss function that distinguishes between changepoints attributable to shifts in local versus global network structure. In particular, we formulate a multinomial log loss function in which a multinomial response encoding changepoint locations is regressed on each network's adjacency matrix and a collection of graph statistics  (e.g., overall edge density, triangle count). These two components distinguish local and global changes, respectively. Next, we propose a prior for the coefficient matrix of the network's adjacency matrix based on the network shrinkage prior of \citet{guha2023high}. This prior combines common low-rank assumptions in the network literature~\citep{hoff2008, athreya2018statistical, loyalchen2023spike} with a spike-and-slab prior~\citep{ishwaran2005spike} to  account for the network's topological structure.  Moreover, the sparsity pattern induced by the proposed prior allows us to identify changepoints due to shifts related to individual dyads or small groups of nodes. Our last contribution is a fast Gibbs sampler for posterior simulation based on P\'olya--Gamma data augmentation~\citep{polson2013bayesian} that is linear in the observed edges in the dynamic network, enabling our method to scale to networks of realistic size.

The rest of the article is organized as follows. Section~\ref{Sec. methods} presents the proposed generalized Bayesian changepoint method for dynamic networks and its implementation. Section~\ref{Sec. Simulations} evaluates the proposed method against competing methods in simulation studies. Section~\ref{Sec. Real data application} applies the method to detecting changepoints in a real email communication network and international conflict network. Section~\ref{Sec. Discussion} concludes with a discussion.

\section{Generalized Bayes for network changepoint analysis}\label{Sec. methods}


\subsection{The generalized Bayesian approach} \label{Sec. bcmlrNET method}


Consider a time series of networks on $n$ nodes observed over $T$ time points represented by a collection of $n \times n$ adjacency matrices $\bm{A}_{1:T} = \{\bm{A}_t\}_{t=1}^T$ with entries $A_{ij,t} \in \mathbb{R}$. We assume each network is undirected with no self-loops, so that each $\bm{A}_t$ is symmetric with zeros on the diagonal. 
We suppose there exist $K-1$ changepoints $\kappa_1, \dots, \kappa_{K-1} \in  \{1, \dots, T-1\}$ satisfying $0 = \kappa_0 < \kappa_1 < \dots < \kappa_{K-1} < \kappa_{K} = T$ that divide the series into $K$ segments. Let $\bm{\kappa} = (\kappa_1, \dots, \kappa_{K-1})^{\top}$ and label the $k$th segment as $\mathcal{S}_k = \left\{\kappa_{k-1} +1, \dots, \kappa_{k}\right\}$ for $1 \leq k \leq K$.

Our objective is to develop a coherent Bayesian procedure for inferring the locations and nature of the changepoints $\bm{\kappa}$ without relying on a particular statistical network model, which is often misspecified in practice. To do so, we adopt the generalized Bayesian framework~\citep{zhang2006, bissiri2016general, martin2022} for updating one's prior beliefs about a parameter $\theta$ that is connected to data $x$ only through a loss function $\ell(\theta; x)$. \citet{bissiri2016general} showed that a coherent update from a prior density $\pi(\theta)$ to a posterior density $\pi(\theta \mid x)$ can be made by setting
$$
\pi(\theta \mid x) \propto \operatorname{exp}\left\{ - \ell(\theta; x)\right\} \pi(\theta),
$$
which does not involve specifying a complete probability model for the data. Moreover, the usual Bayesian update is the special case when $\ell(\theta; x)$ is the negative log likelihood. Note that generalized Bayesian updating has been successfully applied to problems related to changepoint analysis including probabilistic clustering \citep{rigon2023generalized} and sparse high-dimensional classification \citep{mai2026highdimensional}.



In our context, the update from the prior density $\pi\left(\bm{\kappa}, \bm{\theta}\right)$ to the generalized posterior density $\pi\left(\bm{\kappa}, \bm{\theta} \mid \bm{A}_{1:T}\right)$ is
\begin{equation}\label{Eq. unaugmented posterior} 
\pi \left(\bm{\kappa}, \bm{\theta} \mid \bm{A}_{1:T}\right) \propto \operatorname{exp}\left\{ - \ell\left(\bm{\kappa}, \bm{\theta}; \bm{A}_{1:T}\right)\right\} \pi(\bm{\kappa},\bm{\theta}), 
\end{equation}
where $\bm{\kappa}$ denotes the changepoint locations and $\bm{\theta}$ denotes unknown parameters that describe the changepoints, shedding light on the global and local network structures that differ across segments.
Crucially, this generalized Bayesian framework helps address major challenges in changepoint detection in dynamic network data. By replacing the likelihood with a loss function, we avoid the potentially error-prone process of specifying a network-generating process for each segment while still allowing for uncertainty quantification as in a standard Bayesian analysis. As we will see in the following section, a carefully chosen loss function allows for interpretation of changepoints in both local and global network structure.


\subsubsection{The loss function} \label{Sec. bcmlrNET loss function}

The inspiration for our loss function comes from the classification-based approach to changepoint detection in which unknown segment memberships, e.g., $\mathbbm{1}\{t \in \mathcal{S}_k\}$, serve as class labels for a multivariate classifier~\citep{londschien2023, thomas2024bayesian, wang2026bcmlr}. In particular, we take the perspective that a good partition $\bm{\kappa}$ of the network time series is one for which a classifier can reliably distinguish the network structure across adjacent segments, e.g., $\mathcal{S}_k$ and $\mathcal{S}_{k+1}$. This motivates us to define our loss function as the following multinomial log loss:
\begin{align} \label{Eq. bcmlrNET loss}
    \ell\left(\bm{\kappa}, \bm{\theta}; \bm{A}_{1:T}\right)
    &= -\log \left\{\prod_{t=1}^T \prod_{k=1}^K q_{k}(\bm{A}_t)^{1\{t \in \mathcal{S}_k\}}\right\} \nonumber\\ 
    &=-\log\left\{
    \prod_{t=1}^{\kappa_1}q_{1}(\bm{A}_t) \times
    \dots \times
    \prod_{t=\kappa_{K-1}+1}^{T}q_{K}(\bm{A}_t)
    \right\},
\end{align}
where the class probability $q_{k}(\bm{A}_t) = \textrm{Pr}(t \in \mathcal{S}_k \mid \bm{A}_t)$ is the probability, according to the chosen classifier, that time point $t$ is assigned to segment $\mathcal{S}_k$ given the network $\bm{A}_t$. Unlike in the usual multinomial log loss, the changepoint locations $\bm{\kappa}$ are unknown and treated as parameters in the loss function. 


To complete the specification of the loss function, we parameterize the class probabilities in terms of a multivariate classifier that can identify changes due to shifts in both local and global network structure. To do so, we design a classifier that uses information from both the full adjacency matrices $\bm{A}_t$ and a $p$-dimensional vector of network summary statistics $s(\bm{A}_t)$ 
(e.g., edge count, triangle count). For interpretability, we parameterize the class probabilities in terms of the following multinomial logistic regression classifier:
\begin{align} \label{Eq. q_tk and varphi_tk}
    q_{k}(\bm{A}_t) = \frac{\exp\{\varphi_{k}(\bm{A}_t)\}}{\sum_{l=1}^K\exp\{\varphi_{l}(\bm{A}_t)\}},
    \qquad
    \varphi_{k}(\bm{A}_t) =
    \frac{1}{2}\langle \bm{A}_t,\bm{B}_k\rangle_F + 
    s(\bm{A}_t)^{\top}\bm{\beta}_k, 
\end{align}
where $\bm{B}_1, \dots, \bm{B}_K \in \mathbb{R}^{n \times n}$ are symmetric local-level coefficient matrices,  $\bm{\beta}_1, \dots, \bm{\beta}_K \in \mathbb{R}^p$ are global-level coefficient vectors, and $\left\langle \cdot , \cdot \right\rangle_F$ denotes the Frobenius inner product satisfying 
$\left\langle \bm{A}_t,\bm{B}_k \right\rangle_F = \operatorname{tr} \left( \bm{A}_t^{\top}\bm{B}_k \right)$. The linear predictor $\varphi_{k}(\bm{A}_t)$ combines local dyad-level information from the adjacency matrix $\bm{A}_t$ with global information from the summary statistics $s(\bm{A}_t)$ to accurately classify the segments. If the local component $\frac{1}{2}\langle \bm{A}_t,\bm{B}_k\rangle_F = \sum_{i < j} A_{ij,t}B_{ij,k}$ is omitted from  the linear predictor, this loss reduces to the loss used in the changepoint method for vector-valued time series of \citet{wang2026bcmlr}. However, the local component is essential for detecting changes in local network structure that often occur in network time series. Moreover, the generalized posterior can determine whether the local component is necessary for distinguishing the segments on either side of a changepoint,  as demonstrated in Section~\ref{Sec. Simulations} and Section~\ref{Sec. Real data application}. 

Unlike many existing network changepoint methods, the parameterization in \eqref{Eq. q_tk and varphi_tk} leads to interpretable summaries of the causes of the network changepoints.  The coefficients can be interpreted through the multinomial logistic regression model implied by the loss function when the changepoints $\bm{\kappa}$ are known.
For the $k$th changepoint, the log-odds of $\bm{A}_t$ belonging to segment $\mathcal{S}_{k+1}$ rather than segment $\mathcal{S}_k$ is
$$
\log\frac{q_{k+1}(\bm{A}_t)}{q_{k}(\bm{A}_t)} = \frac{1}{2} \left\langle \bm{A}_t,\bm{B}_{k+1}-\bm{B}_k \right\rangle_F + s(\bm{A}_t)^{\top} \left( \bm{\beta}_{k+1}-\bm{\beta}_k \right).
$$
Hence, 
$\bm{B}_{k+1}-\bm{B}_k$ and $\bm{\beta}_{k+1}-\bm{\beta}_k,$ respectively, characterize the local and global network features that distinguish the segments on either side of the changepoint. A positive/negative $(i,j)$th entry of $\bm{B}_{k+1}-\bm{B}_k$ indicates that the presence of an edge between nodes $i$ and $j$ is more/less characteristic of networks in segment $\mathcal{S}_{k+1}$ than $\mathcal{S}_{k}$. Similarly, a positive/negative $q$th entry of $\bm{\beta}_{k+1}-\bm{\beta}_k$ for $q \in \{1,\dots,p\}$ indicates that larger values of the $q$th network summary statistic are more/less characteristic of networks in segment $\mathcal{S}_{k+1}$ than $\mathcal{S}_k$. 

We incorporate network structure into the matrix-valued regression coefficients $\bm{B}_{1:K} = \left\{\bm{B}_k\right\}_{k=1}^K$ by imposing a low-rank constraint commonly found in the literature on network-predictor regression~\citep{arroyo2019, wang2019, guha2021brain, guha2023high}.
For $1 \leq k \leq K$, we constrain $\bm{B}_k$ to have rank less than or equal to $D \ll n$. More precisely, we set
\begin{align}\label{Eq. B's low-rank factorization}
\bm{B}_k = \bm{U}_k\bm{\Lambda}_k\bm{U}_k^{\top},
\end{align}
where $\bm{U}_k=(\bm{u}_{1k},\ldots,\bm{u}_{nk})^{\top}\in\mathbb{R}^{n\times D}$ contains node-specific latent vectors as rows and $\bm{\Lambda}_k=\operatorname{diag}(\lambda_{1k},\ldots,\lambda_{Dk})$ is a $D \times D$ diagonal matrix with $\lambda_{dk}\in\{-1,0,1\}$. The reduced-rank decomposition in \eqref{Eq. B's low-rank factorization} is reminiscent of the generalized random dot product graph (RDPG) model~\citep{athreya2018statistical, rubindelanchy2022}, which is a type of network latent space model~\citep{hoff2002latent, hoff2008}. Accordingly, we interpret the latent vectors $\bm{u}_{1k}, \dots, \bm{u}_{nk}$ as the positions of the nodes in a $D$-dimensional latent space such that their bilinear products determine the strength of the local-level effects, e.g., $B_{ij,k} = \bm{u}_{ik}^{\top}\bm{\Lambda}_k\bm{u}_{jk}$. This bilinear structure implicitly encodes transitive network effects, as discussed in \citet{hoff2005}. Moreover, since each $\bm{\Lambda}_k$ can have zeros on the diagonal, the rank $D_{k} = \sum_{d=1}^D \mathbbm{1}\{\lambda_{dk} \neq 0\} \leq D$ of each $\bm{B}_k$ may be different. Because local changepoints may involve only a small subset of nodes, another advantage of this decomposition is that it will allow us to place sparsity-inducing priors on the node-specific latent positions in order to identify the subset of nodes associated with a detected change.



We omit intercept terms from $\varphi_{k}(\bm{A}_t)$ for $1 \leq k \leq K$ because they are not identifiable and their inclusion can lead to pathological behavior of the posterior distribution of changepoint locations; see Section~S1.1 of the Supplementary Material of \citet{thomas2024bayesian}. 
We center and standardize both $\bm{A}_{1:T}$ and $\bm{s}(\bm{A}_1), \dots, s(\bm{A}_T)$. In particular, for each dyad $\{i,j\}$, we subtract the sample mean of $\{A_{ij,t} \, : \, t = 1, \dots, T\}$ and divide by its sample standard deviation. We apply this same procedure separately to each network summary statistic across time. We refer to the proposed method as \texttt{bcmlrNET}, a stylized acronym for ``generalized Bayesian changepoint detection in networks with a loss inspired by multinomial logistic regression.''

\subsubsection{Prior specification}\label{Sec. priors}

As in multinomial logistic regression, the segment-specific coefficients require a reference class for identifiability.  We take segment $\mathcal{S}_K$ as the reference class and set
$\bm{B}_K$ and $\bm{\beta}_K$ to a matrix and vector of zeros, respectively. The priors for segment-specific parameters below therefore apply to
$k\in\{1,\dots,K-1\}$. 

To induce node-level sparsity, we place spike-and-slab
priors~\citep{ishwaran2005spike} on the node-specific latent
positions, following the network shrinkage construction of
\citet{guha2021brain,guha2023high}. Specifically, for $1 \leq i \leq n$ and $1 \leq k \leq K-1$, we assign $\bm{u}_{ik}$ the prior 
$$
\bm{u}_{ik}\mid z_{ik}, \tau^2 \overset{\text{ind.}}{\sim}
\begin{cases}
\delta_{\bm{0}}, & \text{if } z_{ik}=0,\\
\mathcal{N}(\bm{0},\tau^2\bm{I}_D), & \text{if } z_{ik}=1,
\end{cases}
$$
where the binary variable $z_{ik}\in\{0,1\}$ indicates whether $\bm{u}_{ik}$ is assigned to the spike or slab, and $\delta_{\bm{0}}$ denotes a point mass at the zero vector. The prior distribution for the binary variables is 
$$
z_{ik}\mid w_{ik} \overset{\text{ind.}}{\sim} \operatorname{Bernoulli}(1-w_{ik}), \qquad 1 \leq i \leq n, \ 1 \leq k \leq K-1,
$$
where 
$w_{ik}$ is the spike probability. 
From \eqref{Eq. B's low-rank factorization}, we see that assigning $\bm{u}_{ik}$ to the spike sets all dyad-level coefficients involving node $i$ in segment $\mathcal{S}_k$ to zero. If these coefficients are zero in both $\mathcal{S}_k$ and $\mathcal{S}_{k+1}$, then the $i$th row and column of $\bm{B}_{k+1}-\bm{B}_k$ are zero, so node $i$'s local behavior does not distinguish networks in these segments. 
We complete the node-level spike-and-slab hierarchy by assigning independent priors to the spike probabilities $\{w_{ik}  \, : \, 1 \leq i \leq n, \ 1 \leq k \leq K-1\}$ and the shared slab variance $\tau^2$:
$$
w_{ik}\overset{\text{ind.}}{\sim}\operatorname{Beta}(a_w,b_w),
\qquad
\tau^2\overset{\text{ind.}}{\sim}\operatorname{InverseGamma}(a_\tau,b_\tau),
$$
where $b_\tau$ is the scale parameter. 
We set $a_w=b_w=1$, resulting in a uniform prior for each $w_{ik}$. The choices of $a_\tau$ and $b_\tau$ are guided by the conditional prior variance of the dyad-level coefficients as detailed in Proposition~\ref{prop:prior_variance_B}.


We now specify the prior distribution  of the diagonal elements $\lambda_{1k}, \ldots, \lambda_{Dk}$ of the matrix $\bm{\Lambda}_k$ for each $k \in \{1, \ldots, K-1\}.$ Recall that these parameters determine the rank $D_{k} = \sum_{d=1}^D \mathbbm{1}\{\lambda_{dk} \neq 0\} \leq D$  of the coefficient matrix $\bm{B}_k.$ 
Following \citet{guha2021brain,guha2023high}, we assign a hierarchical prior distribution to $\lambda_{1k}, \ldots, \lambda_{Dk}$ that favors a low-rank coefficient matrix $\bm{B}_k.$ 
We suppose that there exists a probability vector $\bm{\pi}_d = (\pi_{d1}, \pi_{d2}, \pi_{d3})^\top$ for each latent dimension $d \in \{1, \ldots, D\}$ such that the $\lambda_{dk}$ parameters are conditionally independent given $\bm{\pi}_1, \ldots, \bm{\pi}_D$ with 
\begin{align*}
\Pr(\lambda_{dk}=c\mid \bm{\pi}_1, \ldots, \bm{\pi}_D)=
\begin{cases}
\pi_{d1}, & c=0,\\
\pi_{d2}, & c=1,\\
\pi_{d3}, & c=-1.
\end{cases}
\end{align*}
The probability vectors are distributed $\bm{\pi}_{d} \stackrel{\text{ind.}}{\sim} \text{Dirichlet}(d^\eta, 1, 1)$ \textit{a priori} for some $\eta > 1.$ Under this prior specification, $\Pr(\lambda_{dk}=0)=d^\eta/(d^\eta+2)$ increases with $d$ so that latent dimensions with larger indices are more likely to be excluded. The expected rank of $\bm{B}_k$ is
\begin{align} \label{eq:expected_rank}
\operatorname{E}(D_k)=
\sum_{d=1}^D\frac{2}{d^\eta+2},
\end{align}
from which we see that larger $\eta$ values promote lower rank coefficient matrices. We set $\eta=1.01$ following the recommendation of \citet{guha2023high}.

The coefficient $B_{ij,k} = \bm{u}_{ik}^{\top}\bm{\Lambda}_k\bm{u}_{jk}$ determines the contribution of dyad $\{i,j\}$ to the linear predictor $\varphi_{k}(\bm{A}_t)$ in \eqref{Eq. q_tk and varphi_tk}. An overly diffuse prior for $B_{ij,k}$ places substantial mass on dyad-level effects that imply unreasonably large changes in the log odds of segment membership. A similar issue is discussed in the context of the standard logistic regression model in \citet{gelman2008weakly}. It is therefore important to understand how the prior variance of $B_{ij,k}$ depends upon the prior hyperparameters $a_\tau, b_\tau,$ and $\eta$ introduced above.
\begin{proposition}\label{prop:prior_variance_B} Under the prior specification above with $a_\tau > 2$, for any dyad $\{i,j\}$ and non-reference segment $k,$ we have that 
    $$
    \operatorname{E}\left(B_{ij,k} \mid z_{ik}=z_{jk}=1\right)=0,
    $$
    and
    $$
    \operatorname{Var}\left(B_{ij,k} \mid z_{ik}=z_{jk}=1\right) 
    = \operatorname{E}\left(\tau^4\right)\operatorname{E}(D_k), 
    $$
    where $\operatorname{E}(\tau^4)=b_\tau^2/\{(a_\tau-1)(a_\tau-2)\}$ and $\operatorname{E}(D_k)$ is given in \eqref{eq:expected_rank}.
\end{proposition}
\noindent We have already chosen $\eta=1.01$ based on the recommendation of \citet{guha2023high}. We recommend setting $a_\tau = 4$ to ensure that the prior for $\tau^2$ is not extremely heavy-tailed. 
Section E.1 of the Supplementary Material describes how we can use  Proposition~\ref{prop:prior_variance_B} and Chebyshev's inequality to identify a reasonable range of values for $b_\tau$ that can then be further narrowed down through experiments. In the end, we recommend setting $b_\tau = 1/2. $

For the changepoints $\bm{\kappa},$ we adopt the prior proposed by \citet{wang2026bcmlr}:
\begin{align}\label{Eq. kappa prior}
\pi(\bm{\kappa}\mid K) \propto \prod_{k=1}^{K} \left(\frac{1}{\kappa_k-\kappa_{k-1}} \right)^{\kappa_k-\kappa_{k-1}} \times \mathbbm{1}\left\{\min_{1 \leq k \leq K}|\mathcal{S}_k| \geq m\right\}.
\end{align}
This prior mass function downweights changepoint configurations with closely spaced changepoints and is maximized when the changepoints are as evenly spaced as possible. The indicator term imposes an integer minimum segment length $m > 1$ such that $mK \leq T$, which is a common practice in the changepoint literature~\citep{zhao2019, bhattacharjee2020, londschien2023}.


Lastly, to the coefficients of the network summary statistics, we assign independent Gaussian priors, with  $\bm{\beta}_k \sim \mathcal{N}(\bm{m}_{0k},\bm{V}_{0k})$ for $1 \leq k \leq K-1$, where
$\bm{m}_{0k}$ and $\bm{V}_{0k}$ are a mean vector and
positive-definite covariance matrix, respectively. 
Throughout the experiments and data analyses in this article, we set $\bm{m}_{0k}$ equal to a vector of zeros and $\bm{V}_{0k} = 3 \bm{I}_p$ for $1 \leq k \leq K-1$.

\subsection{An efficient Gibbs sampler for posterior computation}\label{Sec. bcmlrNET Gibbs sampler}
We propose an efficient Markov chain Monte Carlo (MCMC) algorithm to simulate from the generalized posterior in \eqref{Eq. unaugmented posterior} under the priors outlined in the previous section. Specifically, we develop a Gibbs sampling scheme that can sample the model parameters in $\mathcal{O}(|E|)$ time, where $|E| = \sum_{t=1}^T \sum _{i < j} \mathbbm{1}\{A_{ij,t} \neq 0\}$ is the number of observed edges in the dynamic network. This is an improvement over the typical complexity 
for discrete-time dynamic network models~\citep{sewell2015latent}, which often scale with the number of dyads in the network, that is, $\mathcal{O}(n^2 T)$. The proposed algorithm is more efficient because real-world networks are often sparse, meaning $|E| \ll n^2 T$.


As presented so far, the generalized posterior leads to non-conjugate full-conditional distributions for the model parameters, which makes developing an efficient Gibbs sampler challenging. However, we can use P\'olya--Gamma data augmentation \citep{polson2013bayesian} to obtain tractable full conditional distributions. To permit P\'olya--Gamma data augmentation, we rewrite the loss function \eqref{Eq. bcmlrNET loss} in a form analogous to the negative log likelihood of a multinomial logistic regression model: 
\begin{align}\label{Eq. alternative bcmlrNET loss}
    \ell\left(\bm{y}, \bm{\theta}; \bm{A}_{1:T}\right) \propto
    -\log\left\{
        \prod_{t=1}^T\prod_{k=1}^K
        \frac{e^{\eta_{k}(\bm{A}_t)y_{tk}}}{e^{\eta_{k}(\bm{A}_t)}+1}
    \right\},
\end{align}
where
\begin{align}
    \eta_{k}(\bm{A}_t) = \varphi_{k}(\bm{A}_t)-c_{k}(\bm{A}_t),
    \qquad
    c_{k}(\bm{A}_t) = \log\sum_{l\neq k}\operatorname{exp}\left\{\varphi_{l}(\bm{A}_t)\right\},
\end{align}
and $y_{tk} = \mathbbm{1}\{t \in \mathcal{S}_k\}$ is an indicator that time point $t$ is in segment $\mathcal{S}_k$. The alternative representation in \eqref{Eq. alternative bcmlrNET loss}, introduced by \citet{held2006bayesian}, enables us to augment the generalized posterior density with P\'olya-Gamma variables $\omega_{tk}\sim\operatorname{PG}(1,0)$ for $1 \leq t \leq T$ and $1 \leq k \leq K$, where $\operatorname{PG}$ denotes the P\'olya--Gamma distribution defined by \citet{polson2013bayesian}.

Starting with this augmented posterior density, we present the full conditional distributions necessary to construct a Gibbs sampler in 
Section~A of the Supplementary Material. In addition, we outline how to sample from them in $\mathcal{O}(|E|)$ time. 

\subsection{Selecting the number of changepoints}\label{Sec. Select num of CPs}

The \texttt{bcmlrNET} method presented in Section~\ref{Sec. bcmlrNET method} assumes the number of changepoints is specified in advance, although this number is often unknown in practice. To estimate and quantify uncertainty related to the number of changepoints, we adopt the classification-based selection approach of \citet{wang2026bcmlr}. The first step of this approach is to fit \texttt{bcmlrNET} with a conservatively large number of changepoints $L_{\mathrm{fitted}} = K_{\mathrm{fitted}} -1$ using the Gibbs sampler in Section~\ref{Sec. bcmlrNET Gibbs sampler}. The second step is to determine how many of the fitted changepoints appear to correspond to true changepoints 
by assessing how well the classifier implied by the multinomial log loss \eqref{Eq. bcmlrNET loss} can distinguish networks on either side of each fitted changepoint. 

To evaluate binary classification around each fitted changepoint and estimate the true number of changepoints $L_{\mathrm{true}}$, we use the area under the receiver operating characteristic curve (AUC) \citep{peterson1954theory}. Because an AUC above $0.5$ indicates that a binary classifier can distinguish between classes better than a random coin flip, we use this criterion to determine whether a fitted changepoint appears to correspond to a true changepoint. We compute the AUC from binary labels and prediction probabilities using the \texttt{pROC} package \citep{robin2011proc}. Specifically, for $k \in \left\{1, \dots, L_{\mathrm{fitted}}\right\}$ at Gibbs iteration $s\in\{1,\dots,S\}$, the binary labels are the indicators $y_{t,k+1}^{(s)}=\mathbbm{1}\{t \in \mathcal{S}_{k+1}^{(s)}\}$, 
and the prediction probability is the conditional probability that network $\bm{A}_t$ belongs to $\mathcal{S}_{k+1}^{(s)}$ given that it belongs to either $\mathcal{S}_{k}^{(s)}$ or $\mathcal{S}_{k+1}^{(s)}$ based on the class probabilities in \eqref{Eq. q_tk and varphi_tk},
$$
\tilde{q}_{k+1}^{(s)}(\bm{A}_t)
=\frac{q_{k+1}^{(s)}(\bm{A}_t)}{
q_k^{(s)}(\bm{A}_t)+q_{k+1}^{(s)}(\bm{A}_t)}, \qquad t \in \mathcal{S}_k^{(s)} \cup \mathcal{S}_{k+1}^{(s)}. 
$$

Let $\mathrm{AUC}_k^{(s)}$ denote the AUC computed from these labels and probabilities. If $\mathrm{AUC}_k^{(s)} >0.5$, then $\kappa_k^{(s)}$ appears to be a true changepoint at iteration $s.$ Similarly, the number of fitted changepoints that appear to be true changepoints at iteration $s$ is $L_{\mathrm{true}}^{(s)} := \sum_{l=1}^{L_{\mathrm{fitted}}}I_l^{(s)}$, where $I_k^{(s)} = \mathbbm{1}\{\mathrm{AUC}_k^{(s)}>0.5\}$. In the end, the samples $L_{\mathrm{true}}^{(1)},\ldots,L_{\mathrm{true}}^{(S)}$ approximate the posterior distribution of the number of fitted changepoints that appear to be true changepoints under the model with $L_{\mathrm{fitted}}$ changepoints. We take the posterior mode of this distribution as our point estimate of $L_{\mathrm{true}}$.

In practice, we implement a slightly modified version of this procedure, as recommended in \citet{wang2026bcmlr}. To reduce the risk of selecting spurious changepoints due to overfitting, we hold out every $\zeta$th network for AUC evaluation and fit the model to the remaining networks. Because each AUC estimate may be based on relatively few held-out networks, we require the lower bound of its $(1-\alpha)\times100\%$ confidence interval, rather than the point estimate, to exceed the threshold $\xi=0.5$. Finally, if $\hat{L}_{\mathrm{true}}>0$, we recommend refitting \texttt{bcmlrNET} to the full series with $\hat{L}_{\mathrm{true}}$ changepoints. This restores all admissible changepoint locations in the original series and facilitates interpretation of the network features that distinguish adjacent segments.
These refinements to the procedure above are described in detail in Section~C 
of the Supplementary Material. Our implementation is available in the \texttt{bcmlrNET} repository\footnote{\url{https://github.com/YuhuiChloe/bcmlrNET}}.
\section{Simulation studies} \label{Sec. Simulations}

In this section, we present simulation studies comparing the proposed method with four competing methods in terms of their ability to detect changepoints that occur in both local and global network structure for dynamic network data generated from several well-known statistical network models. 


We consider the following four competing methods, which differ according to their category (i.e., whether they are model-based, vectorized, metric space, or embedding methods) and the type of changepoint they are designed to detect: \texttt{CPDrdpg}~\citep{madridpadilla2022cpdrdpg}, \texttt{CPDker}~\citep{madridpadilla2021cpdker}, \texttt{CPDnbs}~\citep{wang2021cpdnbs}, and \texttt{CPDstergm}~\citep{kei2023cpdstergm}. The \texttt{CPDrdpg} method is an embedding method that assumes an underlying dynamic random dot product graph model and detects changes in node-level latent position distributions using a nonparametric CUSUM statistic. It can detect changepoints at the node and dyad level. The \texttt{CPDker} method is a vectorized method that detects distributional shifts in a sequence of multivariate network summary statistics. In our simulation studies, these network summary statistics correspond to edge and triangle counts, which describe global network structure. The \texttt{CPDnbs} method is network binary segmentation, a vectorized method that identifies changes by maximizing CUSUM statistics based on vectorized weighted adjacency matrices. It can detect both local and global changes.
Lastly, the \texttt{CPDstergm} method is a model-based method that fits a time-heterogeneous separable temporal exponential-family random graph model \citep{krivitsky2014separable} with group fused lasso regularization to identify changes in formation and dissolution parameters associated with selected network summary statistics.  It targets global-level changes. 

To highlight the advantages of incorporating local information from the adjacency matrices in addition to global information from the network summary statistics, we fit three different versions of the proposed method: 
(1) \texttt{bcmlrNET}\textsubscript{A}, which fixes $\bm{\beta}_1,\dots,  \bm{\beta}_K $ to vectors of zeros so that only the adjacency matrices are available, (2) \texttt{bcmlrNET}\textsubscript{S}, which fixes $\bm{B}_1, \dots, \bm{B}_K $ to matrices of zeros so that only the summary statistics are available, and (3) \texttt{bcmlrNET}\textsubscript{A+S}, the full method which leaves all of these parameters unconstrained so that both sources of information are available. 
Note that \texttt{bcmlrNET}\textsubscript{S} corresponds to the \texttt{bcmlr} method proposed by \citet{wang2026bcmlr}. The first two reduced methods are presented only for the purpose of comparison. Outside of the context of this simulation study, we recommend using the full \texttt{bcmlrNET}\textsubscript{A+S} method, and we drop the $\text{A+S}$ subscript. 
In our simulation studies, we supply \texttt{bcmlrNET}\textsubscript{A+S} and \texttt{bcmlrNET}\textsubscript{S} with edge and triangle counts, which match the summary statistics supplied to the competing \texttt{CPDker} method.

To assess the accuracy of each changepoint method, we use two metrics: the adjusted Rand index (ARI) \citep{morey1984measurement, hubert1985comparing} and the percentage of replicates in which the correct number of changepoints is detected. Given a partition of the data based on the true changepoints and a partition of the data based on estimated changepoints, the Rand index \citep{rand1971objective} is equal to the fraction of the $\binom{T}{2}$ pairs of observations that are either in the same segment in both partitions or in different segments in both partitions. The adjusted Rand index (ARI) is the normalized difference between the Rand index and the expected Rand index for a random partition. An ARI of 1 indicates perfect changepoint estimation. 

A complete description of the implementation details, including the hyperparameter settings for our method and the competing methods, can be found in Section D 
of the Supplementary Material.

\subsection{Simulation scenarios}

We evaluate our method and the competing methods in four scenarios that differ in terms of the types of changepoint (local, global, or both) and the statistical network model used to generate the networks. 
For each scenario, we generate binary symmetric networks with $n = 50$ nodes observed over $T = 300$ time points with two changepoints occurring at $\kappa_1 = 75$ and $\kappa_2 = 225$. 
This process is repeated to produce 50 independent replicates for each scenario. The data-generating processes are described in the following sections. 




\subsubsection{Scenario 1: stochastic block models} \label{Sec. Sim_SBMs}
In this scenario, the network-generating process is based on the stochastic block model (SBM)~\citep{holland1983stochastic}. An SBM assigns each node to a community and assumes that the probability of an edge between two nodes depends only on their community membership. 
Here, we consider two communities 
and define the baseline edge probability $p_{ij,t}$ of dyad $\{i,j\}$ at time $t$ as
$
p_{ij,t} = \bm{z}_{i,t}^{\top} \mathcal{B} \bm{z}_{j,t}, 
$
where $\bm{z}_{i,t}, \bm{z}_{j,t} \in \{(1,0)^{\top}, (0,1)^{\top}\}$ represent node $i$'s and node $j$'s community memberships at time $t$. The block-connectivity matrix $\mathcal{B} = (p - q) \mathbf{I}_2 + q\mathbf{1}_2 \mathbf{1}_2^{\top}$ has a within-community connection probability $p$ and a between-community connection probability $q$.
\begin{figure}[h]
    \centering
    \includegraphics[width=\textwidth]{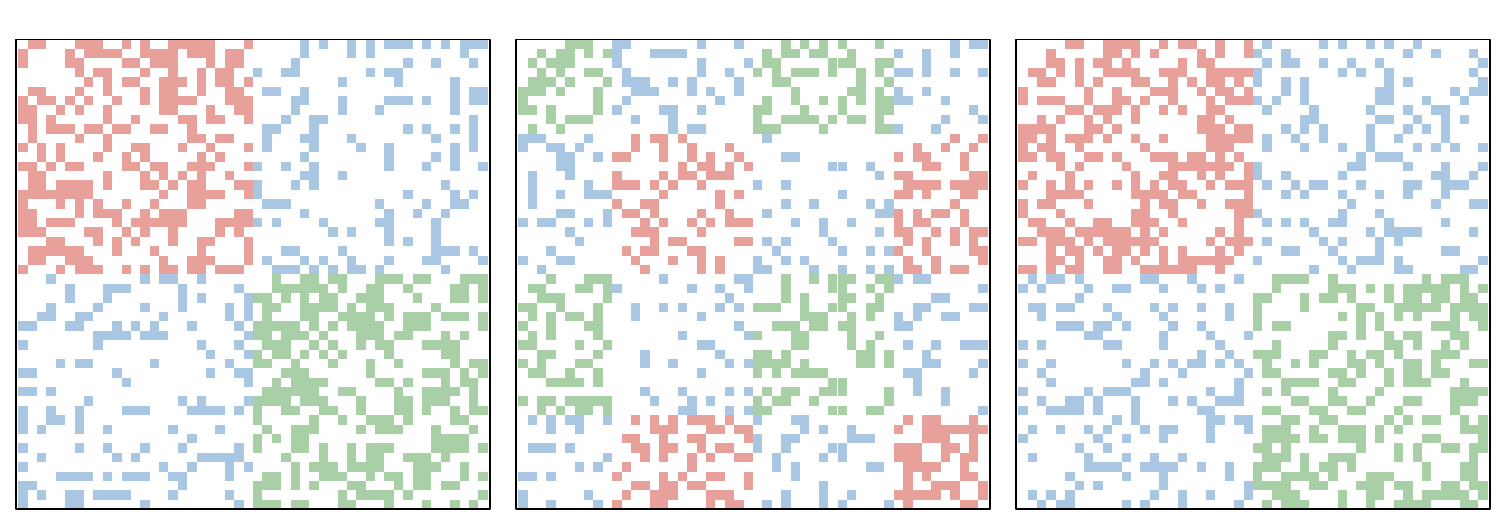}
    \caption{Adjacency matrices from segments $\mathcal{S}_1, \mathcal{S}_2$, and $\mathcal{S}_3$ (left to right) under scenario 1, where the changes arise from exchanges of community membership in a stochastic block model. The edges between two nodes from community 1 are in pale red, the edges between two nodes from community 2 are in pale green, and the edges between two nodes from different communities are in light blue. The changes in the block memberships appear in the adjacency matrices as subtle differences in local edge densities across changepoints.}
    \label{fig: Adjacency matrices under SBMs}
\end{figure}

Following \citet{kei2023cpdstergm}, we introduce temporal dependence by letting the edge probability of each dyad depend on the presence of an edge at the preceding time point. 
Specifically, we sample the dyads 
at time $t = 1$ as
\begin{equation}\label{Eq. Initialize dyads in SBM}
A_{ij,1} = A_{ji,1} \mid p_{ij,1} \overset{\text{ind.}}\sim \operatorname{Bernoulli}\left(p_{ij,1}\right), \quad 1 \leq i < j \leq n,
\end{equation}
and then update the dyads 
at subsequent time indices $t = 2, \ldots, T$ as 
\begin{equation}\label{Eq. Update dyads with temporal dependence}
A_{ij,t} = A_{ji, t} \mid p_{ij,t}, A_{ij,t-1} \overset{\text{ind.}}\sim
\begin{cases}
\operatorname{Bernoulli}\!\left(\rho\left(1-p_{ij,t}\right)+p_{ij,t}\right), & A_{ij,t-1}=1, \\
\operatorname{Bernoulli}\!\left((1-\rho)p_{ij,t}\right), & A_{ij,t-1}=0,
\end{cases}
\end{equation}
where $0 \leq \rho < 1$. For $\rho=0$, there is no temporal dependence, while a value of $\rho$ close to 1 leads to very strong temporal dependence. In this scenario, we set $\rho = 0.1$. 

\begin{figure}[h]
    \centering
    \includegraphics[width=\textwidth]{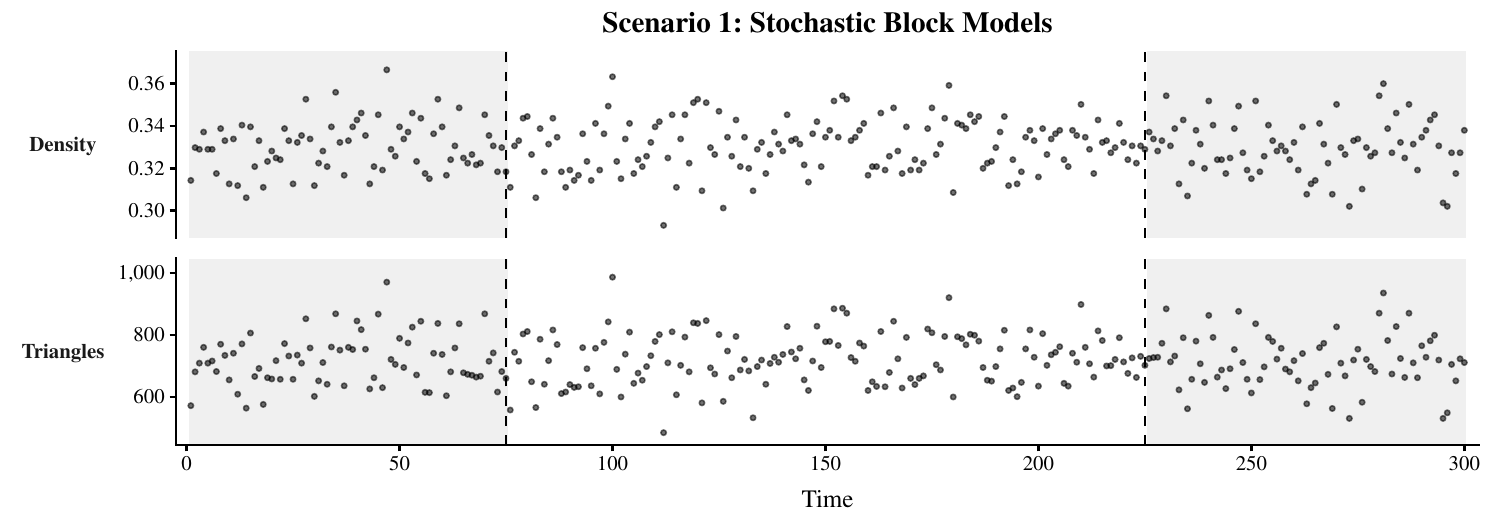}
    \caption{Network densities and triangle counts of a series of adjacency matrices under scenario 1. The stable trends of these statistics show that they are not informative about the changepoints.}
    \label{fig: network stats under SBMs}
\end{figure}

To generate networks in which the changepoints occur at the dyad-level, we allow the community memberships 
to change at the changepoints, 
while keeping the block-connectivity matrix $\mathcal{B}$ fixed. In the first segment, we assign the first 25 nodes to Community 1 and the last 25 nodes to Community 2. After the first changepoint, the first 10 nodes and the last 10 nodes switch communities. After the second changepoint, these nodes return to their original communities. For each of the 50 replicates, we sample $p \sim \operatorname{Uniform}(0.4, 0.5)$ and $ q \sim \operatorname{Uniform}(0.2, 0.3)$ to induce assortative communities.

Figure \ref{fig: Adjacency matrices under SBMs} and Figure \ref{fig: network stats under SBMs} visualize one replicate from the network-generating process described above. Figure \ref{fig: Adjacency matrices under SBMs} displays three adjacency matrices from segments $\mathcal{S}_1, \mathcal{S}_2$, and $\mathcal{S}_3$, respectively. We highlight edges in different colors based on their community memberships. The changes in the block memberships appear in the adjacency matrices as subtle differences in local edge densities across changepoints. Figure \ref{fig: network stats under SBMs} plots the time series of the two summary statistics supplied to \texttt{bcmlrNET}\textsubscript{A+S}, \texttt{bcmlrNET}\textsubscript{S}, and the competing methods, which show no clear changepoint. Thus, in this scenario, we expect only local information to be informative.



\subsubsection{Scenario 2: random dot product graph models}
In this scenario, the network-generating process is based on the random dot product graph (RDPG) \citep{young2007random}.
Under the RDPG, the edge probability $p_{ij,t}$ of dyad $\{i,j\}$ at time $t$ is 
$
p_{ij,t}=\bm{x}_{i,t}^{\top}\bm{x}_{j,t},
$
where $\bm{x}_{i,t}, \bm{x}_{j,t} \in \mathbb{R}^3$ are the latent position vectors of nodes $i$ and $j$. Given the edge probabilities, the edges are conditionally independent Bernoulli random variables with
$
A_{ij,t} \mid p_{ij,t} \stackrel{\text{ind.}}{\sim} \operatorname{Bernoulli}\left(p_{ij,t}\right)
$
for $1 \leq i < j \leq n.$


To generate the changepoints, we assign each node piecewise-constant latent positions.
In the first segment, each node $i$ has a latent position vector $\bm{x}_{i,t} = \bm{x}_{i}^*$ consisting of the first three coordinates of an independent sample from a $\operatorname{Dirichlet}\left(\frac{1}{4},\frac{1}{4},\frac{1}{4},\frac{1}{4}\right)$. 
After the first changepoint, the first 5 nodes and the last 5 nodes experience a shift of position so that
$$
\bm{x}_{i,t}=
\begin{cases}
(1-\delta_x)\bm{x}_i^*+\delta_x\bm{e}_1, & i \in \{1,\dots,5\}, \\
\bm{x}_i^*, & i \in \{6,\dots,n-5\}, \\
(1-\delta_x)\bm{x}_i^*+\delta_x\bm{e}_2, & i \in \{n-4,\dots,n\},
\end{cases}
$$
where $\bm{e}_1=(1,0,0)^\top,\bm{e}_2=(0,1,0)^\top,$ and $\delta_x =0.3$.  After the second changepoint, the latent position vector of each node $i$ reverts to the baseline vector $\bm{x}_{i}^*$. This network-generating process results in changepoints that are evident at the dyad-level, with small changes also appearing at the global-level.

\begin{figure}[t]
    \centering
    \includegraphics[width=\textwidth]{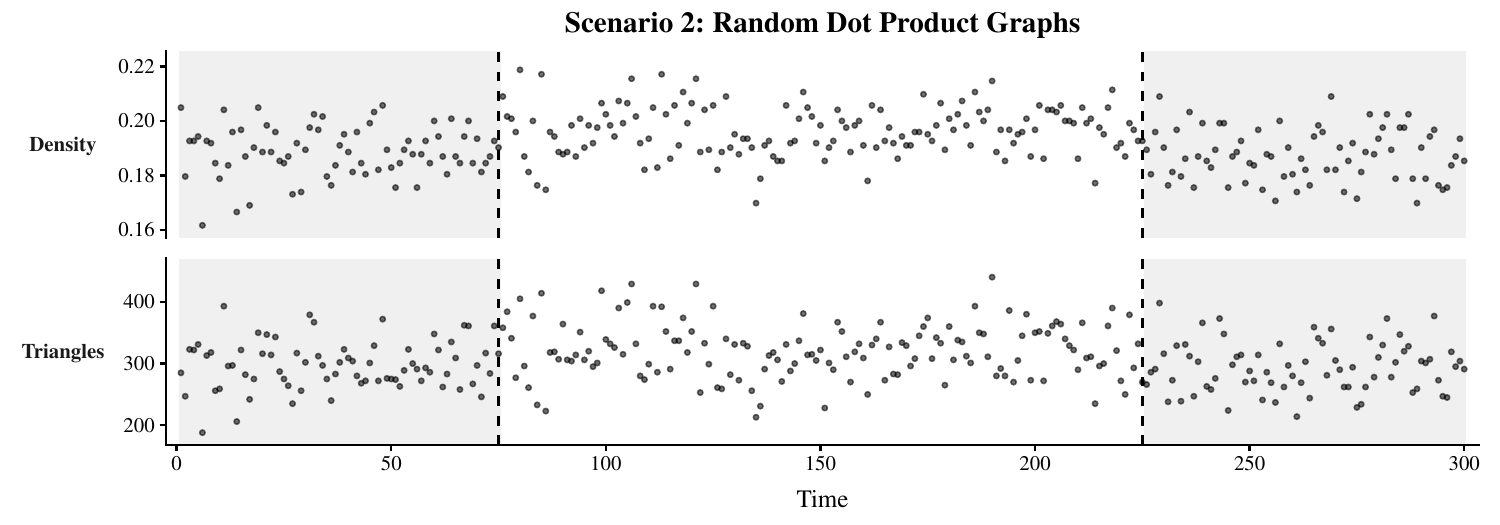}
    \caption{Network density and the number of triangles of a series of adjacency matrices under scenario 2. Both statistics are generally higher in segment $\mathcal{S}_2$ than in segments $\mathcal{S}_1$ and $\mathcal{S}_3$ . }
    \label{fig: network stats under RDPGs}
\end{figure}

Figure \ref{fig: network stats under RDPGs} visualizes the network summary statistics of one replicate from the network-generating process described above. The networks in segment $\mathcal{S}_2$ are slightly denser and have more triangles than those in segments $\mathcal{S}_1$ and $\mathcal{S}_3$, reflecting the modest increase in overall edge probabilities induced by the dyad-level latent position shifts. 

\subsubsection{Scenario 3: exponential-family random graph models}
In this scenario, we generate networks from independent exponential-family random graph models (ERGM) \citep{frank1986markov}. Under an ERGM, the probability of an observed adjacency matrix $\bm{a}_t$ is
$
\Pr(\bm{A}_t=\bm{a}_t \mid \bm{\theta}_t ) \propto
\exp\left\{\bm{\theta}_t^\top\bm{s}(\bm{a}_t)\right\},
$
where $\bm{s} \, : \, \mathbb{R}^{n \times n} \mapsto\mathbb{R}^p$ is a vector of network summary statistics and $\bm{\theta}_t \in \mathbb{R}^p$ is the corresponding coefficient vector. 
\begin{figure}[h]
    \centering
    \includegraphics[width=\textwidth]{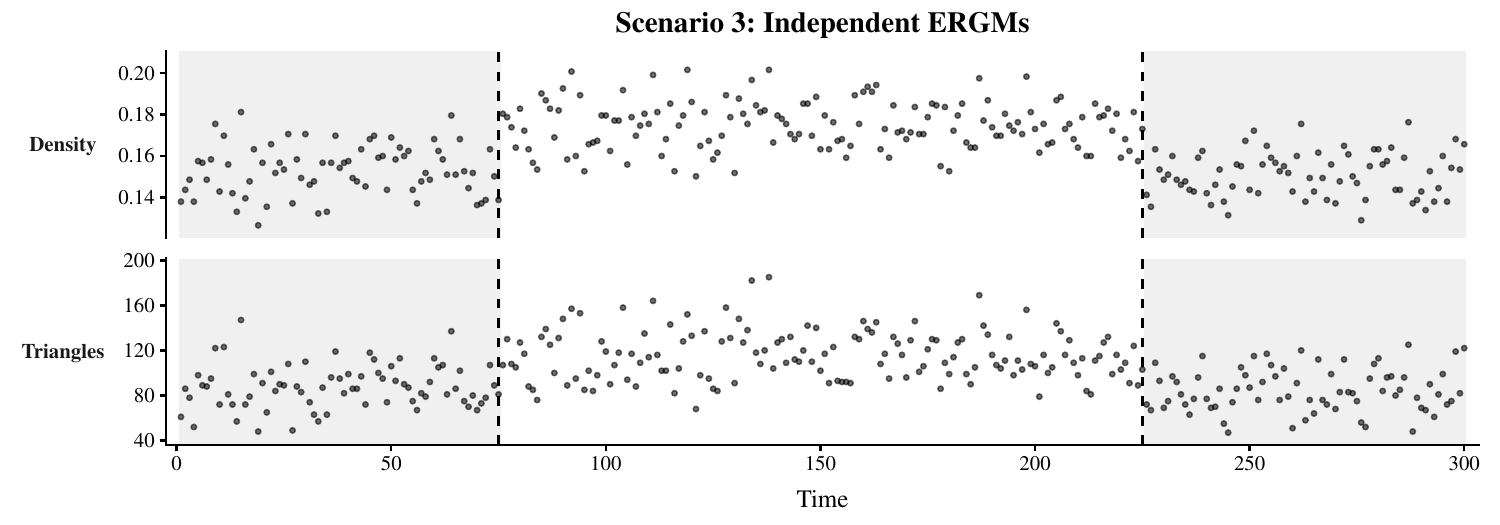}
    \caption{Network densities and triangle counts of a series of adjacency matrices under scenario 3. The network densities and triangle counts increase noticeably in segment $\mathcal{S}_2$ versus segments $\mathcal{S}_1$ and $\mathcal{S}_3$ .}
    \label{fig: network stats under ERGMs}
\end{figure}
Here, we use two network statistics: the number of edges and the geometrically weighted edgewise shared partner (GWESP) statistic with fixed decay parameter 0.5 \citep{snijders2006new,hunter2007curved}. The edge statistic controls the overall tendency for edges to occur, while the GWESP statistic promotes triangle formation.

In this scenario, changepoints only occur at the global level through changes in the coefficients of the network statistics. 
Specifically, in the first segment, $\bm{\theta}_t = \bm{\theta}^*$, where $\bm{\theta}^* = (-2.10,\,0.20)^\top$ is the baseline coefficient vector, with the first and second entries being the coefficients for the edge and GWESP statistics, respectively. After the first changepoint, $\bm{\theta}_t$ changes to $(-1.80,\,0.12)^\top$. After the second changepoint, $\bm{\theta}_t$ reverts to $\bm{\theta}^*$.  Figure \ref{fig: network stats under ERGMs} visualizes one replicate from the network-generating process described above. We observe changes at the global level, with the network densities and triangle counts being visibly higher in segment $\mathcal{S}_2$ than in the other two segments. 


\subsubsection{Scenario 4: Mixed membership stochastic block models}
In this scenario, the network-generating process is based on the mixed-membership stochastic block model (MMSBM) \citep{airoldi2008mixed}. MMSBMs associate each node with a vector of community membership probabilities, rather than assigning each node exclusively to a single community, as in a stochastic block model. We consider four communities and denote the $4\times 4$ block-connectivity matrix at time $t$ as $\mathcal{B}_t.$ The baseline edge probability of dyad $\{i,j\}$ at time $t$ is
$
p_{ij,t} = \bm{\pi}_i^\top \mathcal{B}_t \bm{\pi}_j,
$
where the vectors of community membership probabilities are independently drawn $\bm{\pi}_i \sim \operatorname{Dirichlet}(\bm{\alpha}_i)$ for $i =1, \dots, n$. We set the concentration parameters $\bm{\alpha}_i$ so that nodes 1--10, 11--20, 21--35, and 36--50 have primary memberships in community 1, 2, 3, and 4, respectively. In particular, we let $g_i\in\{1,\dots, 4\}$ denote the primary community of node $i$ and set $\alpha_{ic} = 4 \, \mathbbm{1}\{c = g_i\} + (1/4) \, \mathbbm{1}\{c \neq g_i\}$ for $c = 1, \dots, 4$ so that each node is more likely to belong to its primary community while retaining positive membership probability in the other communities. As in Scenario 1, we introduce temporal  dependence using equations \eqref{Eq. Initialize dyads in SBM} and \eqref{Eq. Update dyads with temporal dependence} with $\rho = 0.1$ to generate the networks.
\begin{figure}[h]
    \centering
    \includegraphics[width=\textwidth]{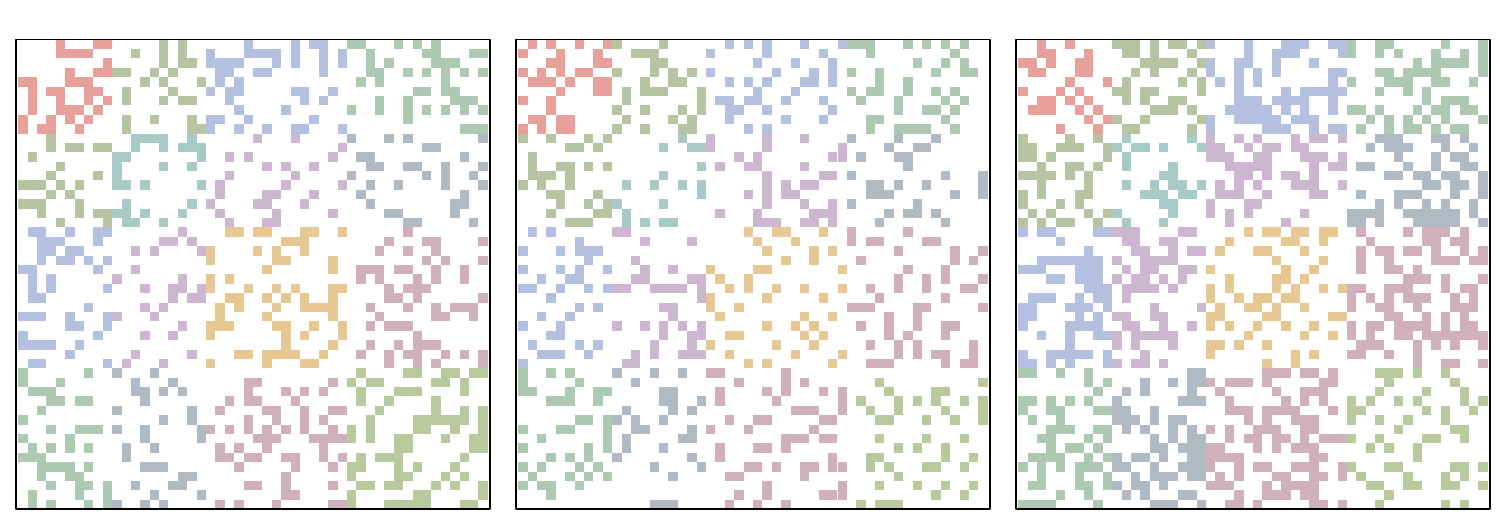}
    \caption{Adjacency matrices from segments $\mathcal{S}_1, \mathcal{S}_2$, and $\mathcal{S}_3$ (left to right) under scenario 4. The changepoints arise from changes in block-connectivity probabilities under a mixed-membership stochastic block model, which are reflected in both subtle differences in local edge densities in the adjacency matrices and changes in network summary statistics.}
    \label{fig: Adjacency matrices under MMSBMs}
\end{figure}


\begin{figure}[htb]
    \centering
    \includegraphics[width=\textwidth]{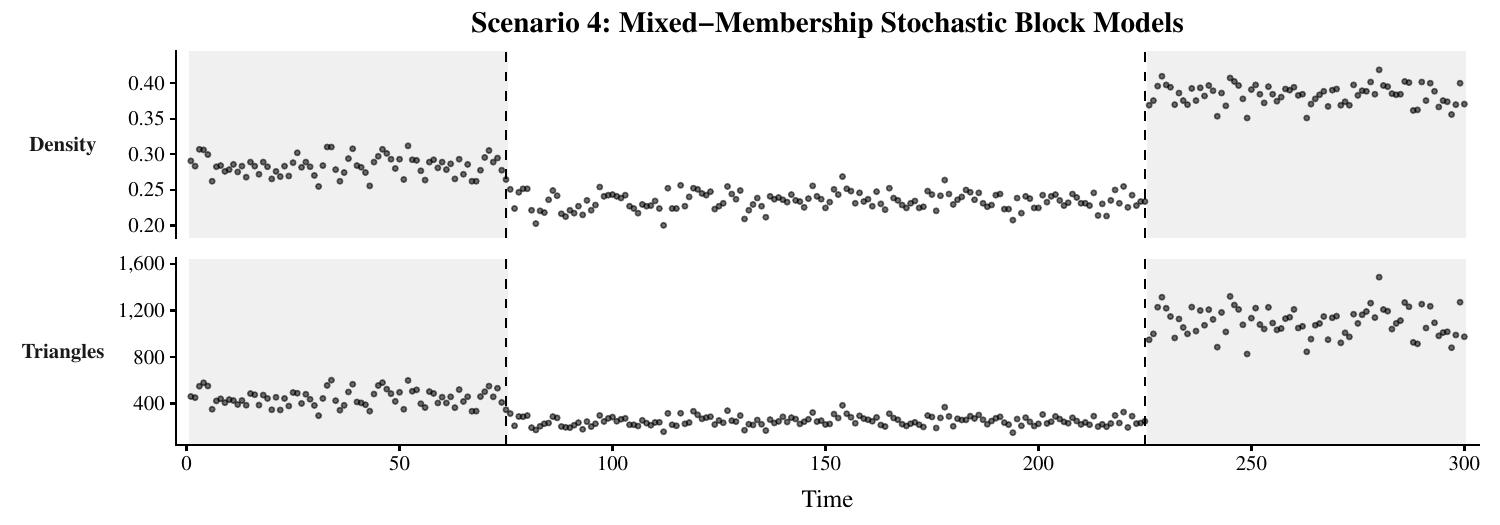}
    \caption{Network densities and triangle counts of a series of adjacency matrices under scenario 4. The network densities and triangle counts show distinct trends of decrease and increase across segments.}
    \label{fig: network stats under MMSBMs}
\end{figure}

We create dyad-level changepoints by changing the block-connectivity matrix $\mathcal{B}_t$, which also creates global-level changes observable in the network summary statistics. In the first segment, $\mathcal{B}_t = (p - q) \mathbf{I}_4 + q\mathbf{1}_4 \mathbf{1}_4^{\top}$ has a within-community connection probability $p$ on the diagonal and a between-community connection probability $q$ on the off-diagonal. After the first changepoint, all diagonal entries decrease by $\delta$. After the second changepoint, all off-diagonal entries increase by $\delta$. In this scenario, we set $\delta = 0.2$ and generate $p \sim \text{Uniform}(0.4, 0.5)$ and $q \sim \text{Uniform}(0.2, 0.3)$ as in Section \ref{Sec. Sim_SBMs}. 

Figure \ref{fig: Adjacency matrices under MMSBMs} and Figure \ref{fig: network stats under MMSBMs} visualize one replicate from the network-generating process in this scenario. Figure \ref{fig: Adjacency matrices under MMSBMs} displays three adjacency matrices from segments $\mathcal{S}_1, \mathcal{S}_2$, and $\mathcal{S}_3$, respectively. We highlight edges in different colors based on their community memberships. The changes in block-connectivity probabilities appear as differences of local edge densities across segments. Figure \ref{fig: network stats under MMSBMs} plots the time series of the two summary statistics, which show distinct trends of decrease and increase across segments.

\subsection{Simulation results}\label{Sec. Simulation results}

Table \ref{tab:simulation_ari_correct_cps} presents simulation results across 50 replicates for each scenario and method. The subscripts A, S, and A+S attached to the methods correspond to the types of input data supplied to the method, with A indicating adjacency matrices, S indicating network summary statistics, and A+S indicating both adjacency matrices and network summary statistics. The subscripts attached to the reported mean ARIs give the sample standard deviations of the ARIs across the 50 replicates, while the parentheses provide the percentages of replicates in which the correct number of changepoints was detected. 

\begin{table}[H]
\centering
\small
\setlength{\tabcolsep}{5pt}
\renewcommand{\arraystretch}{1.20}
\begin{tabular}{lcccc}
\toprule
Method & SBMs & RDPGs & ERGMs & MMSBMs \\
\midrule
\texttt{bcmlrNET}\textsubscript{A+S} & $0.922_{0.153}\,(76\%)$ & $0.969_{0.149}\,(96\%)$ & $\mathbf{0.972}_{0.106}\,(98\%)$ & $\mathbf{1.000}_{0.002}\,(100\%)$ \\
\texttt{bcmlrNET}\textsubscript{A} & $\mathbf{0.961}_{0.099}\,(86\%)$ & $0.969_{0.149}\,(96\%)$ & $\underline{0.030}_{0.081}\,(0\%)$ & $0.950_{0.144}\,(88\%)$ \\
\texttt{bcmlrNET}\textsubscript{S} & $\underline{0.035}_{0.087}\,(0\%)$ & $0.716_{0.328}\,(66\%)$ & $0.970_{0.068}\,(94\%)$ & $0.993_{0.033}\,(96\%)$ \\
\texttt{CPDrdpg}\textsubscript{A} & $0.058_{0.137}\,(2\%)$ & $\underline{0.052}_{0.144}\,(0\%)$ & $0.319_{0.393}\,(26\%)$ & $0.960_{0.063}\,(100\%)$ \\
\texttt{CPDker}\textsubscript{S} & $0.263_{0.160}\,(2\%)$ & $0.391_{0.196}\,(12\%)$ & $0.637_{0.169}\,(32\%)$ & $0.506_{0.119}\,(2\%)$ \\
\texttt{CPDnbs}\textsubscript{A} & $0.414_{0.030}\,(0\%)$ & $\mathbf{0.978}_{0.003}\,(100\%)$ & $0.623_{0.231}\,(60\%)$ & $\underline{0.407}_{0.031}\,(0\%)$ \\
\texttt{CPDstergm}\textsubscript{A} & $0.377_{0.140}\,(4\%)$ & $0.620_{0.284}\,(6\%)$ & $0.816_{0.171}\,(14\%)$ & $0.920_{0.141}\,(66\%)$ \\
\bottomrule
\end{tabular}
\caption{
Mean ARIs across 50 simulation replicates; subscripts give the sample standard deviations across replicates, and parentheses give the percentages of replicates in which the correct number of changepoints was detected. The highest and lowest mean ARIs within each scenario are boldfaced and underlined, respectively. 
}
\label{tab:simulation_ari_correct_cps}
\end{table}

Comparing \texttt{bcmlrNET}\textsubscript{A+S} to the four competing methods (excluding \texttt{bcmlrNET}\textsubscript{A} and \texttt{bcmlrNET}\textsubscript{S} for now), we see that \texttt{bcmlrNET}\textsubscript{A+S} performs best in terms of mean ARI in three of four scenarios (SBMs, ERGMs, and MMSBMs) and is a close second place after \texttt{CPDnbs}\textsubscript{A} in the remaining scenario (RDPGs). The outcome is similar for the other performance metric, the percentage of replicates in which the correct number of changepoints was detected. These results demonstrate that \texttt{bcmlrNET}\textsubscript{A+S} can accurately detect changes across a variety of different network-generating processes featuring both global and local changes. This flexibility can reasonably be attributed to the generalized Bayes approach, which does not require us to assume the data come from a specific statistical network model producing a specific type of change. 

Next, we compare \texttt{bcmlrNET}\textsubscript{A+S} to \texttt{bcmlrNET}\textsubscript{A} and \texttt{bcmlrNET}\textsubscript{S}. We do not typically know in advance whether network changepoints will be local in nature (evident in the adjacency matrices), global in nature (evident in the summary statistics), or both. Thus, in practice, we recommend the \texttt{bcmlrNET}\textsubscript{A+S} method, which takes in both the adjacency matrices and the summary statistics. Ideally, we would see that \texttt{bcmlrNET}\textsubscript{A+S} achieves accuracy metrics nearly as good as or better than both \texttt{bcmlrNET}\textsubscript{A} and \texttt{bcmlrNET}\textsubscript{S} in all scenarios. That is exactly what we see in Table~\ref{tab:simulation_ari_correct_cps}. The \texttt{bcmlrNET}\textsubscript{A+S} method matches or outperforms both \texttt{bcmlrNET}\textsubscript{A} and \texttt{bcmlrNET}\textsubscript{S} in every scenario except for Scenario 1 (SBMs), in which \texttt{bcmlrNET}\textsubscript{A} is best among all methods and \texttt{bcmlrNET}\textsubscript{A+S} is a close second place. It is not surprising that \texttt{bcmlrNET}\textsubscript{A} does well in this scenario, because the changepoints were designed to be strictly local in nature. Table~\ref{tab:simulation_ari_correct_cps} also shows that \texttt{bcmlrNET}\textsubscript{A} does poorly when changepoints are strictly global (Scenario 3: ERGMs), while \texttt{bcmlrNET}\textsubscript{S} does poorly when changepoints are strictly local (Scenario 1: SBMs), as expected.


Among the competing methods, we see a few surprisingly poor performances in Table~\ref{tab:simulation_ari_correct_cps}. In particular, \texttt{CPDrdpg}\textsubscript{A} does worse than all other methods in Scenario 2, even though the network-generating process is based on the RDPG. The reason for this poor performance is that \texttt{CPDrdpg}\textsubscript{A} consistently underestimates the number of changepoints. Indeed, we can see that it does not choose the correct number of changepoints for any replicates in Scenario 2. On the other hand, the reason for the poor performance of \texttt{CPDnbs}\textsubscript{A} in Scenarios 1 and 4 is that it consistently overestimates the number of changepoints, never choosing the correct number. These tendencies carry over to the analysis of international conflict networks in Section~\ref{Sec. POLECAT}, where we observe that \texttt{CPDrdpg}\textsubscript{A} identifies a single changepoint while \texttt{CPDnbs}\textsubscript{A} identifies 12. 

\section{Real data applications} \label{Sec. Real data application}

We illustrate the practical value of the proposed method through two applications to dynamic network data. In the first application, we examine email communication networks from the collapse of Enron. In the second, we analyze conflict networks between countries. In both applications, we apply \texttt{bcmlrNET} using three network summary statistics: the number of edges, triangles, and isolated nodes. Beyond estimating the number and locations of changepoints, posterior summaries from \texttt{bcmlrNET} allow us to understand the network features associated with detected changes and to quantify the corresponding uncertainty. In both applications, networks are constructed based on non-overlapping weekly intervals, which we label by their start and end dates in the format mm/dd/yyyy--mm/dd/yyyy. Implementation details for \texttt{bcmlrNET} and the competing methods in these applications are provided in Section~D 
of the Supplementary Material.

\subsection{Enron email networks} \label{Sec. Enron}

Enron Corporation was a Houston-based natural gas pipeline company that grew to become a major player in the energy, commodities, and services sectors. Its collapse is one of the most prominent corporate scandals in U.S. history, involving deceptive accounting practices that concealed substantial debt and losses as well as market-manipulation tactics during the 2000--2001 California electricity crisis. In the aftermath of the collapse, the Federal Energy Regulatory Commission (FERC) publicly released a trove of email communications between Enron employees that they had obtained in an investigation.







The Enron Corpus can be viewed as dynamic network data. In particular, \citet{kei2025decoder} constructed a network time series of length $T= 100$ by splitting the corpus into non-overlapping weekly windows, with the first window beginning on 06/05/2000 and the last window ending on 05/05/2002, and defining a network for each week. The nodes in the weekly networks correspond to the 100 most active email accounts throughout the entire time period. An edge exists between two nodes in a weekly network if at least one email was exchanged between the relevant email accounts during that week. 

Our pre-processing of the data differs from that of \citet{kei2025decoder} with respect to one detail. In \citet{kei2025decoder}, the nodes correspond to the 100 most active email \textit{accounts}, but these accounts correspond to only 89 unique \textit{employees}. In contrast, we construct weekly networks with these $n=89$ employees as nodes. We noticed that one employee can have multiple email aliases when we initially fit \texttt{bcmlrNET} to the 100-node network series and found, by examining posterior summaries, that one of the estimated changepoints corresponded to an employee switching from one email alias to another. This changepoint does not tell us anything interesting about the collapse of Enron, but it does highlight the ability of \texttt{bcmlrNET} to identify changes in local network structure. 

\subsubsection{Detected changepoints from \texttt{bcmlrNET} and competing methods}

\begin{figure}[h]
    \centering
    \includegraphics[width=\textwidth]{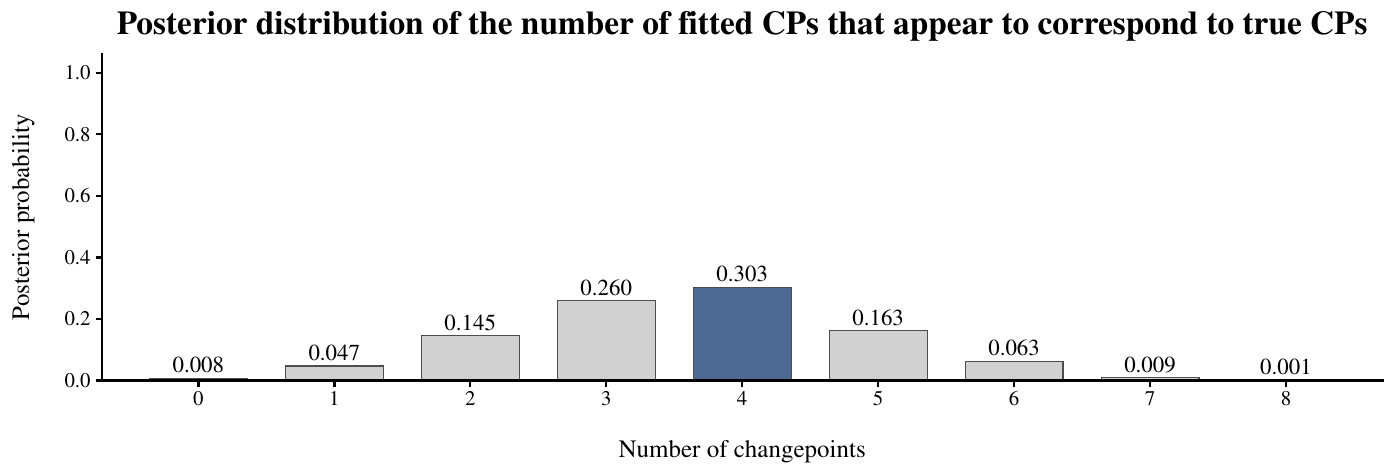}
    \caption{The posterior distribution of the number of fitted changepoints that appear to correspond to true changepoints based on the Enron email network series.}
    \label{fig:enron-numCP-dist}
\end{figure}

\begin{figure}[h]
    \centering
    \includegraphics[width=\textwidth]{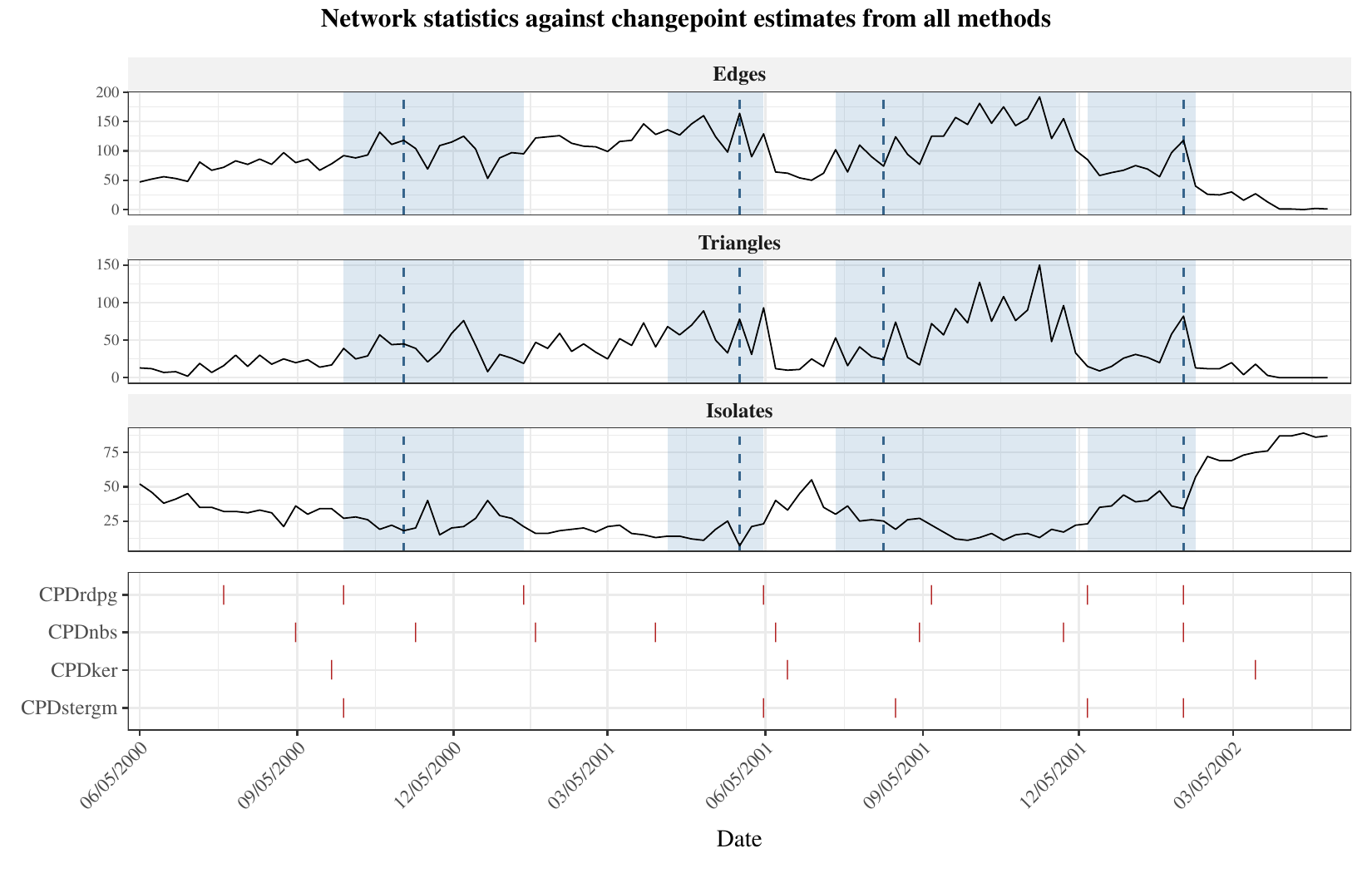}
    \caption{Visualization of detected changepoints plotted over top of the edge, triangle, and isolate statistics calculated from the Enron email network series. The blue dashed lines show the point estimates of changepoint locations from \texttt{bcmlrNET}, while the blue shaded bands show the corresponding 95\% credible intervals. The bottom panel displays estimates of changepoint locations from competing methods.}
    \label{fig:enron_stats_CPs}
\end{figure}

To estimate the number of changepoints and quantify the associated uncertainty,  we apply the approach in Section~\ref{Sec. Select num of CPs} to obtain a posterior distribution over the number of fitted changepoints that appear to correspond to true changepoints (Figure~\ref{fig:enron-numCP-dist}). 
We see considerable uncertainty related to the number of changepoints, which is not surprising given the disagreement among competing methods (Figure~\ref{fig:enron_stats_CPs}). 
As described in Section~\ref{Sec. Select num of CPs}, we take the posterior mode, which we can see is equal to four in Figure~\ref{fig:enron-numCP-dist}, as our point estimate of the number of changepoints and condition on this value for the remainder of the analysis. In light of the uncertainty, we could repeat the analysis for each plausible number of changepoints but, for the sake of brevity, we focus on the case of four changepoints.


Applying \texttt{bcmlrNET} to the entire series with the number of changepoints fixed at four, we arrive at the following point estimates for the changepoint locations: 11/06/2000--11/12/2000, 05/21/2001--05/27/2001, 08/13/2001--08/19/2001, and 02/04/2002--02/10/2002. The first changepoint immediately precedes the filing of Enron's third-quarter Form 10-Q with the Securities and Exchange Commission (SEC) on 11/14/2000. The second changepoint immediately precedes the implementation of the FERC's price-mitigation plan for California's wholesale electricity market on 05/29/2001 \citep[p.~7]{ferc2003californiareview}. The third changepoint coincides with CEO Jeffrey Skilling's abrupt resignation on 08/14/2001 \citep{anderson2002enron}, as well as VP of Corporate Development Sherron Watkins's anonymous letter to Chairman Kenneth Lay on 08/15/2001, which warned that Enron's accounting practices would lead to a major scandal. The fourth changepoint aligns with a series of congressional hearings into the collapse of Enron: Skilling testified and former CFO Andrew Fastow appeared before Congress on 02/07/2002, Lay appeared on 02/12/2002, and Watkins testified on 02/14/2002 \citep{anderson2002enron}. 


Figure~\ref{fig:enron_stats_CPs} plots these point estimates of the changepoint locations as well as 95\% credible intervals over top of the global network summary statistics. The bottom panel allows for comparison with the competing methods. The point estimates from \texttt{bcmlrNET} do broadly align with those found by the other methods, but we still see considerable disagreement across the methods, both in terms of the number and locations of the changepoints, suggesting that there may be considerable uncertainty that goes unreported by the competing methods. These observations highlight the importance of uncertainty quantification in changepoint analysis and confirm a major advantage of \texttt{bcmlrNET}.

\subsubsection{Additional posterior inferences using \texttt{bcmlrNET}}\label{Sec. extra inferences, Enron}



Posterior summaries from \texttt{bcmlrNET} can help us 
interpret the detected changepoints, whether they occur at the local or global level, and quantify the associated uncertainty. In this section, we demonstrate how this can be done in practice. 


\begin{figure}[h]
    \centering
    \includegraphics[width=1.05\textwidth]{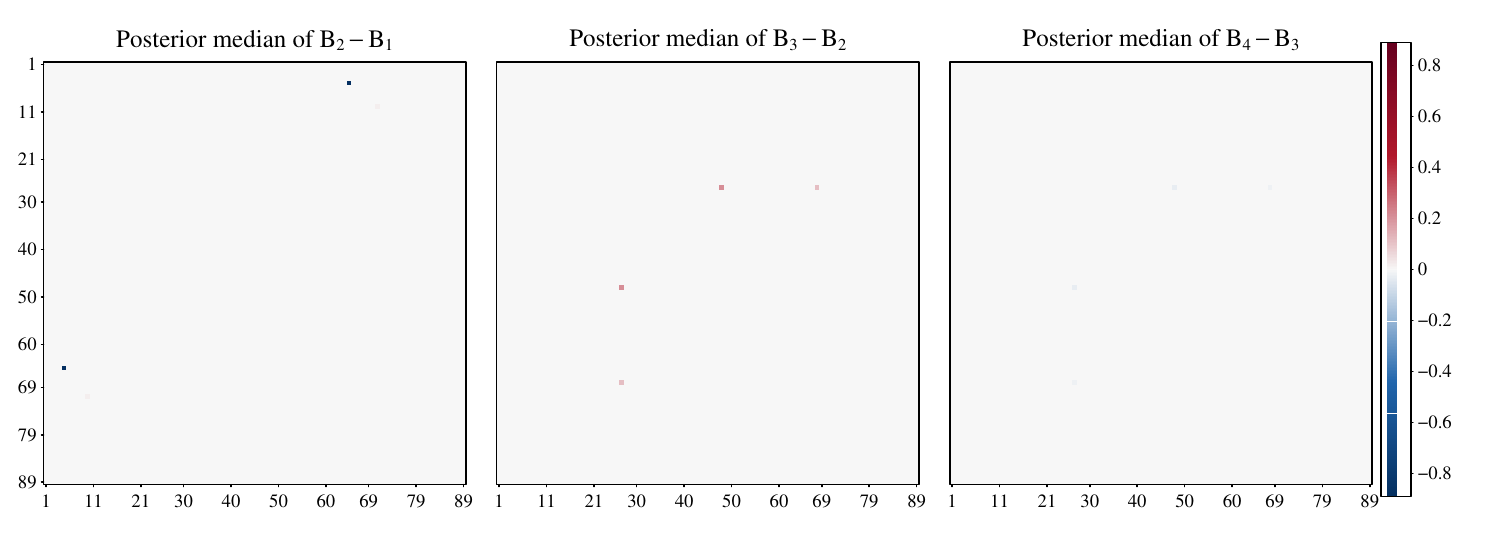}
    \caption{The three heatmaps show the posterior medians of the dyad-level coefficient differences $\bm{B}_2-\bm{B}_1$, $\bm{B}_3-\bm{B}_2$, and $\bm{B}_4-\bm{B}_3$ based on the Enron email network series. The posterior median of $\bm{B}_5-\bm{B}_4$ contains no nonzero entries and is therefore omitted. The common color bar shows the range of dyad-level coefficient differences across all panels, with red and blue indicating positive and negative differences, respectively.}
    \label{fig:enron-B-diff}
\end{figure}

To illustrate interpretation at the dyad level, we focus on the second changepoint, which immediately precedes the implementation of FERC's price-mitigation plan in California's wholesale electricity market. The nonzero entries of the (entry-wise) posterior median of $\bm{B}_3-\bm{B}_2$ (shown as colored entries in the second heatmap in Figure~\ref{fig:enron-B-diff}) correspond to employees connected to Enron's West Power electricity-trading operation. Bill Williams, a trader  on the Real Time Desk within the West Power operation, was directly involved in Enron's electricity trading during the California energy crisis; earlier in 2001, he was recorded asking a power-plant operator to find a reason to take a plant offline during a period of severe electricity crisis (thereby creating artificial scarcity to drive up electricity prices) \citep[pp.~797--798]{nix2021enron}. Kate Symes, a trader on Williams's Real Time trading desk who worked on California imbalance transactions, and Cara Semperger, a senior cash analyst in Enron's West Power operation \citep[Tables~9 and~10]{creamer2022discovering}, also appear among the relevant dyads. The concentration of these employees among the identified dyads provides a substantive connection between the second changepoint and Enron's electricity-trading operations as federal regulators intervened in the California market. The overwhelming majority of dyad-level coefficient differences have posterior medians equal to zero due to the sparsity-inducing priors described in Section~\ref{Sec. methods}, allowing the relatively small set of relevant dyads to stand out. 
Section~D.1 of the Supplementary Material maps the colored dyads in Figure~\ref{fig:enron-B-diff} to Enron employees and reports the signs of their posterior median coefficient differences.

\begin{figure}[ht]
\centering
\begin{subfigure}[t]{0.38\textwidth}
\vspace{0pt}
\centering
\textbf{(a)}\par\vspace{1mm}
\includegraphics[width=\linewidth]{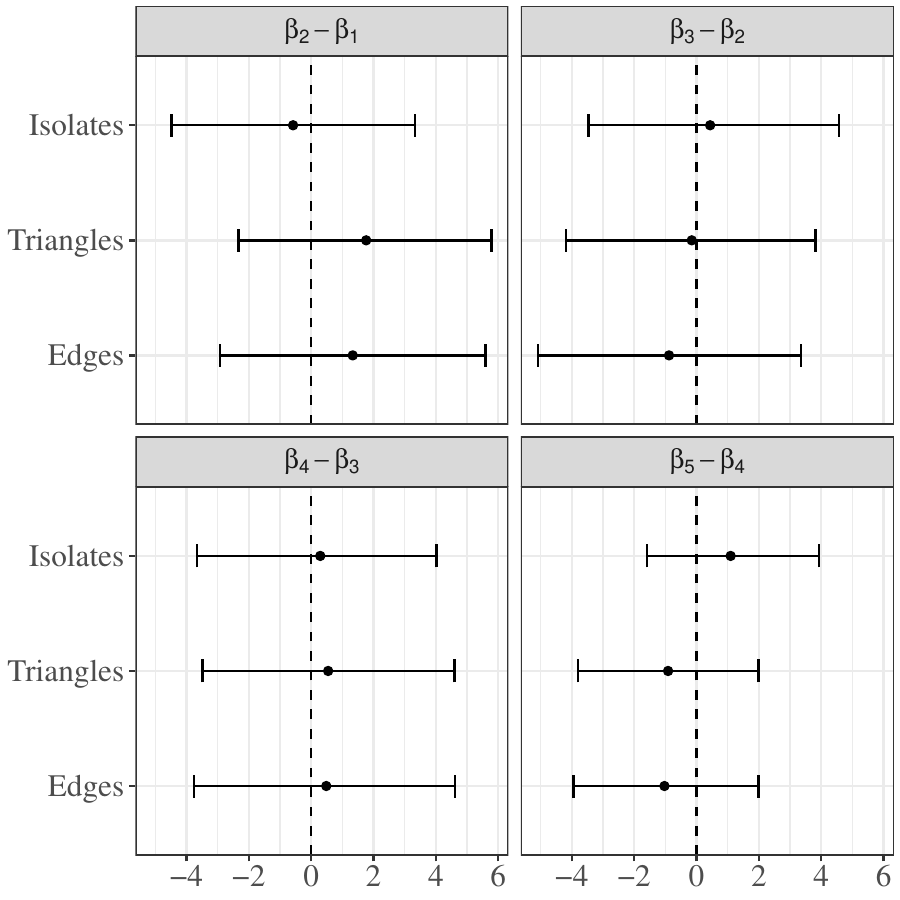}
\label{fig:enron-beta-diff}
\end{subfigure}
\hfill
\begin{subfigure}[t]{0.61\textwidth}
\vspace{-2pt}
\centering
\textbf{(b)}\par\vspace{1mm}
\includegraphics[width=\linewidth]{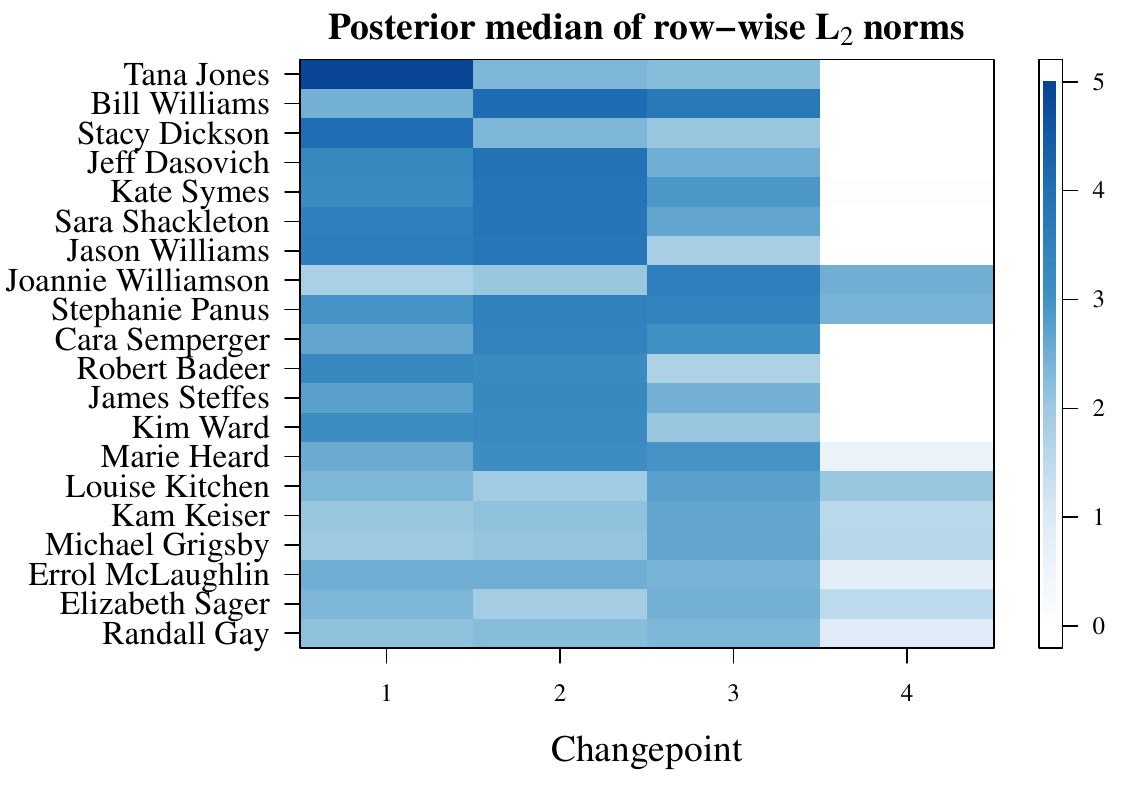}
\label{fig:enron-node-l2}
\end{subfigure}
\captionsetup{skip=4pt}
\caption{
Posterior summaries for interpreting global- and node-level changes in the Enron email network series.
(a) Posterior means (black dots) and 95\% credible intervals for differences in the coefficients of the network summary statistics from adjacent segments.
(b) Union of top eight nodes at each changepoint, ranked by the posterior medians of the node-level $L_2$ norms, e.g., $\left\|\bm{B}_{i,k+1}-\bm{B}_{i,k}\right\|_2$. 
}
\label{fig:enron-glocal-node-changes}
\end{figure}

To illustrate interpretation at the global level, we examine differences in the network statistic coefficients across changepoints, shown in Figure~\ref{fig:enron-glocal-node-changes}(a). At the first changepoint, the estimates suggest increased internal communication before Enron's third-quarter SEC filing deadline. In particular, the estimates for the edge- and triangle-count coefficient differences are positive, while the estimate for the isolate-count coefficient difference is negative. At the fourth changepoint, the signs of the estimates suggest reduced internal communication as Enron faced more intense scrutiny during the congressional hearings. More specifically, the estimates for the edge- and triangle-count coefficient differences are negative, while that for the isolate-count coefficient difference is positive.

To identify the nodes (employees) most involved in detected changepoints, we compute the row-wise $L_2$ norms of differences between coefficient matrices. In particular, for node $i$ at the $k$th changepoint, we compute $
\left\|\bm{B}_{i,k+1}-\bm{B}_{i,k}\right\|_2,
$ which aggregates the magnitude of changes across all dyads involving node $i$. For each changepoint, we rank the top eight nodes according to the posterior medians of these $L_2$ norms. The names in Figure~\ref{fig:enron-glocal-node-changes}(b) come from taking the union of these lists. Bill Williams, who was prominently involved in Enron's energy market manipulation during the California energy crisis, has a large estimated $L_2$ norm at the second changepoint, further connecting this changepoint to the implementation of FERC's price-mitigation plan. For the fourth changepoint, most of the displayed $L_2$ norms are zero or nearly zero. We interpret this observation, along with the earlier findings regarding $\bm{B}_5 - \bm{B}_4$ and $\bm{\beta}_5 - \bm{\beta}_4,$ as further evidence that the fourth changepoint can be attributed to a global reduction in email communications amidst congressional scrutiny rather than a local change in communication patterns.

    \subsection{International conflict networks} \label{Sec. POLECAT}


The period from late 2020 through mid-2024 saw numerous international conflicts, including Russia's full-scale invasion of Ukraine in February 2022 and the outbreak of the Israel-Hamas war in October 2023. News reports of international conflicts can be understood as dynamic network data, with countries as nodes and their conflicts as edges that vary over time. In this section, we apply \texttt{bcmlrNET} to a time series of international conflict networks constructed by \citet{loyal2025generalized} from the Political Event Classification, Attributes, and Types (POLECAT) dataset, which includes records of political events automatically extracted from international news reports \citep{Halterman2023PLOVER, Halterman2023BagOfTricks}. The series consists of $T=184$ non-overlapping weekly networks, starting with the week 12/21/2020--12/27/2020 and ending with the week 06/24/2024--06/30/2024. Each weekly network includes the same $n=197$ countries as nodes. Two countries are connected by an undirected edge if at least one material conflict event was recorded between them during the relevant week. Material conflict events include actions involving physical force, coercion short of violence, military or police mobilization, collective protest, and sanctions or other reductions in existing relations \citep{openeventdataPLOVERmanual}.

\begin{figure}[h]
    \centering
    \includegraphics[width=\textwidth]{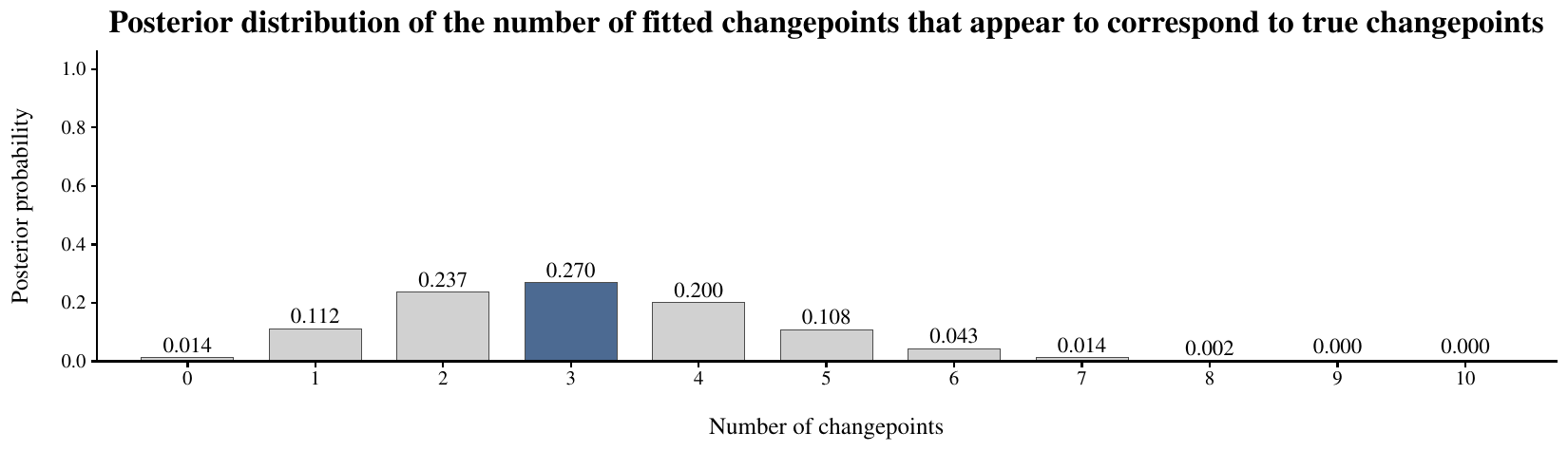}
    \caption{The posterior distribution of the number of fitted changepoints that appear to correspond to true changepoints based on the POLECAT material-conflict network series.}
    \label{fig:polecat-numCP-dist}
\end{figure}

\subsubsection{Detected changepoints from \texttt{bcmlrNET} and competing methods} \label{Sec. detected CPs in POLECAT}
\begin{figure}[h]
    \centering
    \includegraphics[width=\textwidth]{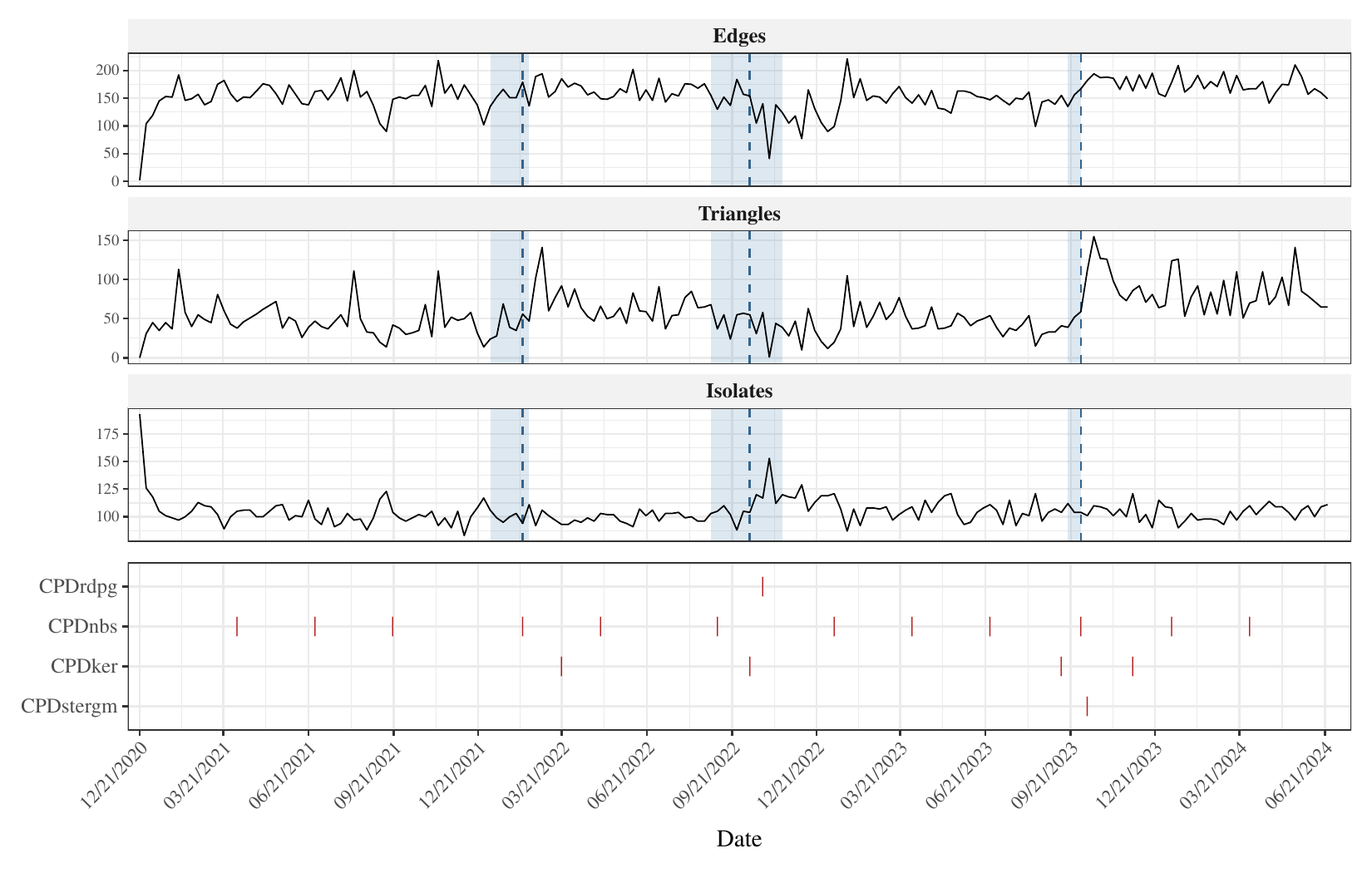}
    \caption{Visualization of detected changepoints plotted over top of the edge, triangle, and isolate statistics calculated from the POLECAT network series. The blue dashed lines denote point estimates of  changepoint locations from \texttt{bcmlrNET}, while the blue shaded bands denote 95\% credible intervals. The bottom panel displays estimates of changepoint locations from competing methods.}
    \label{fig:polecat_stats_CPs}
\end{figure}

To determine the number of changepoints, we once again apply the approach described in Section 2.3. Figure~\ref{fig:polecat-numCP-dist} shows the resulting posterior distribution of the number of fitted changepoints that appear to correspond to true changepoints. As in the Enron application, there is considerable uncertainty related to the number of changepoints. We take the posterior mode of three as our point estimate of the number of changepoints and, for the sake of brevity, we condition on this value for the remainder of the analysis. However, it may be reasonable to repeat the analysis that follows for other plausible numbers of changepoints.

Fitting \texttt{bcmlrNET} to the full series with the number of changepoints fixed at three yields the following point estimates of the changepoint locations: 02/07/2022--02/13/2022, 10/10/2022--10/16/2022, and 10/02/2023--10/08/2023. 
The first changepoint occurs during the military escalation preceding Russia's full-scale invasion of Ukraine on 02/24/2022. On 02/07/2022, the North Atlantic Treaty Organization (NATO) reported that Russia had deployed more than 100,000 troops in and around Ukraine and that approximately 30,000 Russian troops were expected in Belarus \citep{nato2022polandbuildup}. Three days later, NATO reported that Russian troop numbers were continuing to increase \citep{nato2022ukbuildup}. 
The second changepoint coincides with Russia's large-scale missile attacks on cities across Ukraine on 10/10/2022 \citep{un2022ukrainemissiles}. The third changepoint falls in a week with two major conflict developments. On 10/07/2023, Hamas and other Palestinian armed groups launched attacks into Israel, 
followed by Israeli military operations in the Gaza Strip \citep{un2023hamasattack,ocha2023gazaupdate1}. Earlier that week, on 10/05/2023, a missile struck a funeral reception in Hroza, Ukraine. A subsequent United Nations investigation found reasonable grounds to attribute the attack to Russian armed forces \citep{ohchr2023hroza}. 


Figure~\ref{fig:polecat_stats_CPs} plots these point estimates of the changepoint locations, as well as 95\% credible intervals, and allows for comparison with the competing methods. Compared to the Enron application, there is less agreement on the number and locations of changepoints across methods. In particular, \texttt{CPDrdpg} detects one changepoint, while \texttt{CPDnbs} detects twelve. The methods do largely align regarding the third changepoint detected by \texttt{bcmlrNET}, which is identified by four of the five methods. Overall, the results in Figure~\ref{fig:polecat_stats_CPs} once again highlight the high degree of uncertainty that commonly exists in changepoint analysis and reaffirm the importance of methods that offer uncertainty quantification. 

\subsubsection{Additional posterior inferences using \texttt{bcmlrNET}}\label{Sec. extra inferences, POLECAT}



\begin{figure}[h]
    \centering
    \includegraphics[width=0.7\textwidth]{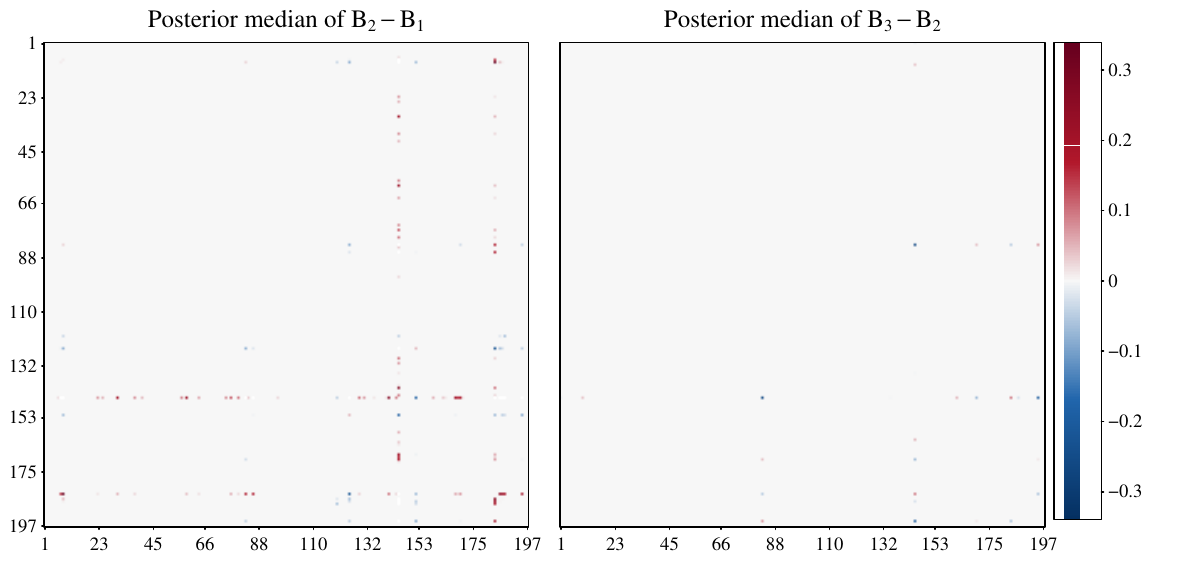}
    \caption{The two heatmaps show the posterior medians of the dyad-level coefficient differences, $\bm{B}_2-\bm{B}_1$ and $\bm{B}_3-\bm{B}_2$, based on the POLECAT network series. The posterior median of $\bm{B}_4-\bm{B}_3$ contains no nonzero entries and is therefore omitted. The common color bar shows the range of dyad-level coefficient differences across all panels, with red and blue indicating positive and negative differences, respectively.}
    \label{fig:POLECAT-B-diff}
\end{figure}

\begin{figure}[ht]
\centering
\begin{subfigure}[t]{0.4\textwidth}
\vspace{0pt}
\centering
\textbf{(a)}\par\vspace{1mm}
\includegraphics[width=\linewidth]{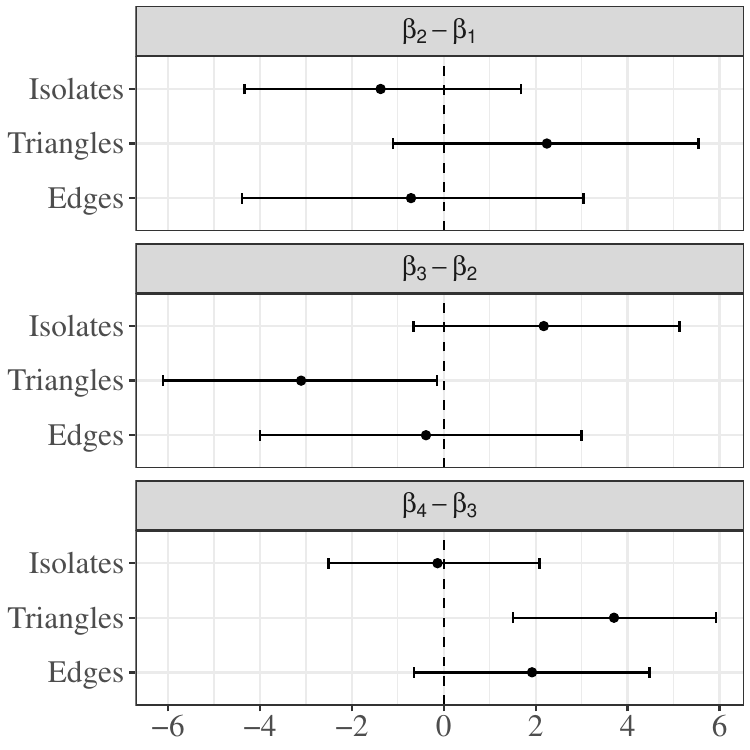}
\end{subfigure}
\hfill
\begin{subfigure}[t]{0.59\textwidth}
\vspace{-2pt}
\centering
\textbf{(b)}\par\vspace{1mm}
\includegraphics[width=\linewidth]{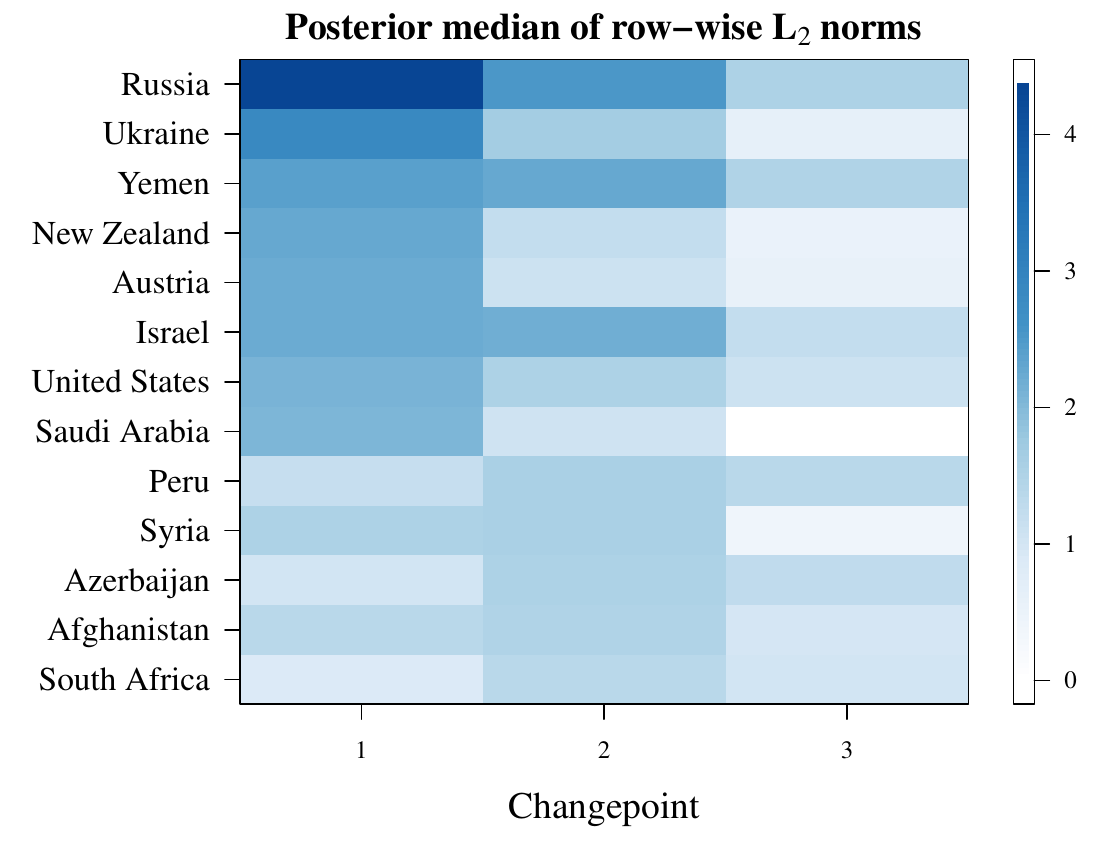}
\end{subfigure}
\captionsetup{skip=4pt}
\caption{
Posterior summaries for interpreting global- and node-level changes in the POLECAT network series.
(a) Posterior means (black dots) and 95\% credible intervals for differences in the coefficients of the network summary statistics from adjacent segments.
(b) Union of top eight nodes at each changepoint, ranked by the posterior medians of the node-level $L_2$ norms, e.g., $\left\|\bm{B}_{i,k+1}-\bm{B}_{i,k}\right\|_2$. 
}
\label{fig:POLECAT-glocal-node-changes}
\end{figure}

As in the Enron application, we can leverage posterior summaries from \texttt{bcmlrNET} to interpret changepoints detected in the international conflict network series. The coefficient difference matrices $\bm{B}_2-\bm{B}_1,$ $\bm{B}_3-\bm{B}_2,$ and $\bm{B}_4-\bm{B}_3$ can help us understand dyad-level changes that may have occurred at the first, second, and third changepoints, respectively. The heatmaps in Figure~\ref{fig:POLECAT-B-diff} display the (entry-wise) posterior medians of $\bm{B}_2-\bm{B}_1$ and $\bm{B}_3-\bm{B}_2.$ All entries of the posterior median of $\bm{B}_4-\bm{B}_3$ are equal to zero, suggesting that the third changepoint corresponds to a global rather than a dyad-level change. Looking at the heatmap corresponding to the posterior median of $\bm{B}_2-\bm{B}_1,$ we see that Russia (node 145) and Ukraine (node 184) are associated with more non-zero dyads (shown as colored entries) than are any other countries. Furthermore, most of the coefficient difference estimates for these dyads are positive (shown as red), indicating an increase in conflict events connected to these dyads at the first changepoint. These observations are consistent with the pre-invasion escalation discussed in Section~\ref{Sec. detected CPs in POLECAT}.

Analogously, the coefficient differences $\bm{\beta}_2-\bm{\beta}_1,$ $\bm{\beta}_3-\bm{\beta}_2,$ and $\bm{\beta}_4-\bm{\beta}_3$ can help us understand global changes in network statistics that may have occurred at each changepoint. 
Notice in Figure~\ref{fig:polecat_stats_CPs} that there is a sharp increase in the triangle count around the third changepoint. This increase is reflected in the third panel of Figure~\ref{fig:POLECAT-glocal-node-changes}(a), which shows that the 95\% credible interval for the triangle-count coefficient difference lies entirely above zero. This indicates that larger triangle counts are more characteristic of networks in segment $\mathcal{S}_4$ than $\mathcal{S}_3$. 
Together with our previous observation that all entries of the posterior median of $\bm{B}_4-\bm{B}_3$ are zero, this observation supports an interpretation in which the third changepoint is driven by a global change in network structure.

To identify the countries most involved in local changes, we use the node-level $L_2$ norms described in Section~\ref{Sec. extra inferences, Enron}. 
Figure~\ref{fig:POLECAT-glocal-node-changes}(b) displays the union of the top eight countries at each changepoint, ranked by the posterior medians of these $L_2$ norms. Russia ranks highest at the first changepoint, with Ukraine also among the most prominent countries, reinforcing the dyad-level interpretation. Israel also appears among the displayed countries, connecting these summaries to another of the conflicts discussed above. The $L_2$ norms associated with the third changepoint are, overall, smaller than those associated with the previous two changepoints. 
This provides additional evidence that the third changepoint is characterized by a global change rather than a local change. 

    \section{Discussion}\label{Sec. Discussion}

The main contribution of this work is a generalized Bayesian approach for identifying the number, locations, and nature of changepoints in dynamic network data. Unlike many existing methods, the proposed method provides interpretable summaries of the causes of changepoints due to shifts in both local and global network structure along with meaningful uncertainty quantification. Moreover, we introduced a fast Gibbs sampler that is linear in the number of observed edges in the network time series, unlike many other Bayesian approaches. We demonstrated the improved performance of our method over competitors on simulated data from a variety of statistical network models. Lastly, we used the proposed method to detect and interpret changepoints in time series of communication and international conflict networks.

There are many directions for future research. The proposed methodology can be extended to directed networks by replacing the low-rank symmetric dyad-level coefficient matrices with an asymmetric representation, i.e., $\bm{B}_k = \bm{U}_k\bm{S}_k\bm{V}_k^{\top}$ where $\bm{U}_k , \bm{V}_k \in \mathbb{R}^{n \times D}$ and $\bm{S}_k$ is a $D  \times D$ diagonal matrix with non-negative entries on the diagonal. In addition, the network summary statistics can be modified to include lagged values, e.g., $s(\bm{A}_t, \bm{A}_{t-1})$, to allow the method to detect changes in network dynamics, e.g., edge persistence, as opposed to just changes in the cross-sectional network structure. Also, network time series often have nodes that enter and leave over time, which is currently not supported by the proposed method due to the size of $\bm{B}_{1:K}$ being fixed. However, variably sized networks can be supported by zeroing out the rows and columns of $\bm{B}_{1:K}$ associated with inactive nodes. Lastly, it would be interesting to extend the proposed methodology to detect changepoints in more complicated network time series, such as dynamic multilayer or bipartite networks. We leave these extensions to future work.

\section*{Availability of data and materials}

Data and code used in the analyses presented in this article are available in the \texttt{bcmlrNET} repository at the following URL: \url{https://github.com/YuhuiChloe/bcmlrNET}. Supplementary Material is available at the following URL: \url{https://michaeljauch.github.io/SupplementbcmlrNETPaper.pdf}.

\section*{Competing interests}
The authors declare no competing interests.
\section*{Funding}
Joshua Daniel Loyal was partially supported by NSF Grant SES-2345043 and NIH Grant UM1TR005128.
\section*{Authors' contributions}
YW: Conceptualization; Methodology; Software; Validation; Visualization; Writing – original draft. MJ: Conceptualization; Methodology; Supervision; Writing – review \& editing. JDL: Conceptualization; Methodology; Supervision; Writing – review \& editing.

\section*{Acknowledgements}
Not applicable.

\putbib

\end{bibunit}

\clearpage

\begin{bibunit}
    \appendix
    \numberwithin{equation}{section}
    \numberwithin{table}{section}
    
\end{bibunit}
\end{document}